\documentclass[sigconf,nonacm]{acmart}

\AtBeginDocument{%
  }

\copyrightyear{2026}
\acmYear{2026}
\acmDOI{}

\acmConference[Conference acronym 'XX]{Make sure to enter the correct
  conference title from your rights confirmation email}{June 03--05,
  2018}{Woodstock, NY}

\acmISBN{}

 \usepackage{multirow} 
 \usepackage{rotating} 
 \usepackage{colortbl}
\usepackage{enumitem}
\usepackage{booktabs}
\usepackage{tabularx}
\usepackage{array}
\usepackage[table]{xcolor}
\usepackage{threeparttable}
\newcolumntype{Y}{>{\raggedright\arraybackslash}X}
\newcolumntype{L}[1]{>{\raggedright\arraybackslash}p{#1}}

\usepackage{pifont}

\usepackage{tabularx}
\usepackage[table]{xcolor}
\usepackage{array}
\usepackage{booktabs}
\usepackage{threeparttable}
\usepackage{tikz}

\definecolor{basic}{HTML}{4E79A7}   
\definecolor{trans}{HTML}{59A14F}   
\definecolor{threeR}{HTML}{E28E2C}  
\definecolor{reg}{HTML}{A06AB4}     

\newcommand{\filledcircle}[1]{%
  \tikz[baseline=-0.55ex]
  \node[
    circle,
    fill=#1,
    draw=#1!70!black,
    line width=0.45pt,
    minimum size=7.5pt,
    inner sep=0pt
  ] {};
}

\newcommand{\hollowcircle}[1]{%
  \tikz[baseline=-0.55ex]
  \node[
    circle,
    fill=white,
    draw=#1!80!black,
    line width=0.60pt,
    minimum size=7.5pt,
    inner sep=0pt
  ] {};
}

\newcommand{\dB}{\filledcircle{basic}}
\newcommand{\oB}{\hollowcircle{basic}}

\newcommand{\dT}{\filledcircle{trans}}
\newcommand{\oT}{\hollowcircle{trans}}

\newcommand{\dR}{\filledcircle{threeR}}
\newcommand{\oR}{\hollowcircle{threeR}}

\newcommand{\dE}{\filledcircle{reg}}
\newcommand{\oE}{\hollowcircle{reg}}

\newcommand{\slotsep}{\hspace{3pt}}

\newcommand{\Bcell}[2]{%
  #1\slotsep#2%
}

\newcommand{\Tcell}[5]{%
  #1\slotsep#2\slotsep#3\slotsep#4\slotsep#5%
}

\newcommand{\Rcell}[4]{%
  #1\slotsep#2\slotsep#3\slotsep#4%
}

\newcommand{\Ecell}[2]{%
  #1\slotsep#2%
}

\newcommand{\stage}[1]{%
  \multicolumn{5}{@{}l@{}}{%
    \cellcolor{gray!12}%
    \hspace{5pt}\textbf{#1}%
  }\\
}
  
\begin{document}

\title{Making Sense of Animal-to-Human Drug Development Evidence: Stakeholder Practices, Challenges, and Requirements for AI Tools}


\author{Rosni Vasu}
\authornote{Rosni Vasu and Simona E. Doneva contributed equally to this research.}
\correspondingauthor
\email{rosni.kottekulam@unibe.ch}
\affiliation{%
  \institution{University of Bern}
  \city{Bern}
  \country{Switzerland}
}

\author{Simona E. Doneva}
\authornotemark[1]
\correspondingauthor
\email{simona.doneva@unibe.ch}
\affiliation{%
  \institution{University of Bern}
  \city{Bern}
  \country{Switzerland}
}

\author{Benjamin V. Ineichen}
\email{benjamin.ineichen@unibe.ch}
\affiliation{%
  \institution{University of Bern}
  \city{Bern}
  \country{Switzerland}
}

\renewcommand{\shortauthors}{Vasu et al.}

\begin{abstract}
Animal models are widely used to study human biology and health interventions, yet translating findings from animal studies to humans remains challenging. Evidence across preclinical and clinical research informs experimental and translational decisions. Artificial Intelligence (AI) tools are increasingly reshaping how this evidence is searched, synthesized, and used, but it remains unclear how they should support the diverse stakeholders involved in assessing animal-to-human evidence. We conducted semi-structured interviews with 13 stakeholders to examine their evidence practices, challenges, and expectations for AI support. We found that stakeholders approach the same incomplete evidence base with different goals, expertise, and heuristics. Participants valued AI particularly for locating, screening, and extracting evidence, but were more cautious about automated interpretation and quality judgments. They emphasized transparency, source traceability, uncertainty communication, and human oversight. Based on these findings, we derive design implications for role-sensitive AI tools that support more systematic and transparent reasoning about animal-to-human translation.

\end{abstract}

\ccsdesc[500]{Human-centered computing~User studies}
\ccsdesc[500]{Human-centered computing~Empirical studies in HCI}
\ccsdesc[500]{Applied computing~Life and medical sciences}

\begin{teaserfigure}
 \includegraphics[width=\textwidth]{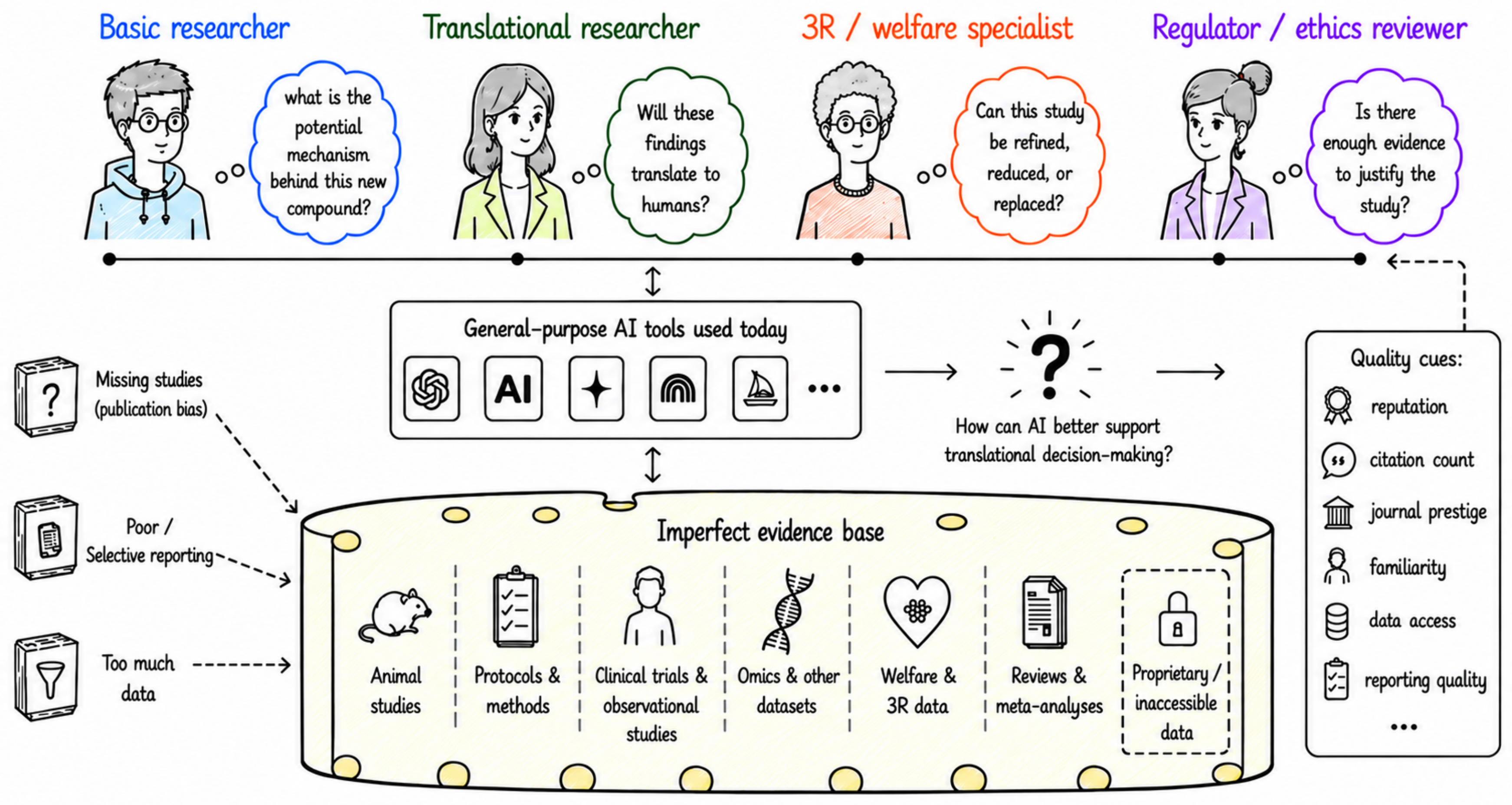}
  \caption{Stakeholders approach the same imperfect evidence base with different goals and quality considerations. General-purpose AI tools can support evidence access and organization, but future systems must better support transparent, context-sensitive translational decision-making.}
\Description{
Four stakeholder groups use the same incomplete evidence base for different decisions. The evidence spans animal and human studies, methods, datasets, reviews, and restricted sources, but is affected by missing results, poor reporting, and information overload. General-purpose AI supports access to this evidence, motivating transparent and context-sensitive support.
}
  \label{fig:teaser}
\end{teaserfigure}

\received{20 February 2007}
\received[revised]{12 March 2009}
\received[accepted]{5 June 2009}

\maketitle


\section{Introduction}

Drug development is a long, costly, and uncertain process \cite{singhDrugDiscoveryDevelopment2023, schlanderHowMuchDoes2021}. Decisions made during animal research -- including which compounds, animal models, and experimental designs to pursue -- can influence whether promising findings progress successfully to clinical trials \cite{donevaLargeScaleAssessmentAnimaltoHuman2026, hogueAssociationStatisticalMethodology2025, zeissEstablishedPatternsAnimal2017}. Yet more than 90\% of drug candidates entering clinical development ultimately fail \cite{ineichenAnalysisAnimaltohumanTranslation2024, kimFactorsAffectingSuccess2023a}. Thus, there is a need to strengthen how evidence from animal studies is generated, assessed, and used in translational research -- i.e., the translation of findings from animals into tangible applications for humans \cite{bespalovFailedTrialsCentral2016a}.

Engaging with prior research can help scientists identify knowledge gaps and design more informative experiments in animals and subsequent human trials \cite{ferreiraLevellingTranslationalGap2020}. Evidence is also used by ethics committees and other oversight bodies to assess whether proposed animal experiments are scientifically justified \cite{wurbelMore3RsImportance2017} and whether they follow the 3Rs: \textbf{R}eplacing animal experiments where possible, \textbf{R}educing animal use, and \textbf{R}efining procedures to minimize distress and improve animal welfare \cite{russell1959principles}. Evidence synthesis practices such as systematic literature reviews, which systematically identify, assess, and integrate findings across relevant studies, have been promoted to strengthen research quality and advance the 3Rs \cite{ineichenSystematicReviewMetaanalysis2024a,devriesUsefulnessSystematicReviews2014b}.

However, evidence synthesis in this domain is particularly challenging because there is an increasing volume of studies available, often differing substantially in their experimental design and reporting practices, while relevant evidence may span partly unpublished sources \cite{ioannidisSystematicReviewsBasic2023, ineichen2023data}. AI-assisted tools may help with tasks such as literature search, screening, information extraction, and evidence organization, and some of these applications have been evaluated in animal-research settings \cite{sousaLandscapeArtificialIntelligence2026,wangPICOEntityExtraction2022a}. Recent advances in language and reasoning models may further support more complex forms of literature analysis \cite{choeSupportingNoviceResearchers2024}. Nevertheless, it remains unclear how such systems should support the diverse goals, responsibilities, and interpretive practices of stakeholders involved in animal and translational research. 

To investigate this question, we conducted semi-structured interviews with 13 stakeholders involved in animal basic and translational research, animal welfare, as well as ethical and regulatory oversight. Participants reflected on their literature practices and challenges and discussed when and how AI tools might support their work. We also elicited feedback on a curated evidence dashboard as a concrete example of AI-assisted evidence support. We were guided by the following research questions (RQs):

\begin{itemize}
    \item[\textbf{RQ1:}] How do researchers and related stakeholders search, evaluate, and synthesize biomedical literature in basic animal and/or translational research? What challenges do stakeholders face in this context?

    \item[\textbf{RQ2:}] What information needs and design requirements emerge for automation / AI tools that support literature analysis and translational evidence assessment?

    \item[\textbf{RQ3:}] How do stakeholders perceive the usefulness and trustworthiness of a curated evidence dashboard?
\end{itemize}

Our findings reveal how stakeholders work with animal and translational evidence, the challenges that constrain this work, and the implications for designing AI-assisted evidence tools.
We make three contributions to HCI research on AI-assisted evidence work:

\begin{itemize}
    \item We provide an empirical account of how stakeholders across basic and translational research, animal welfare and the 3Rs, and regulatory and ethics roles work with animal-to-human evidence. We show that their practices and judgments vary by professional role and decision context, rather than following a single shared process.

    \item We identify 15 evidence-work challenges spanning access, appraisal, study comparison, welfare assessment, cross-species synthesis, and AI use. These challenges show that evidence work is constrained not only by literature volume, but also by absent, inaccessible, and incompletely reported evidence, and that quality assessment requires context-specific expert judgment rather than fixed criteria alone.

    \item We derive design requirements for AI-assisted evidence tools from participants' practices, challenges, and evaluation of a curated evidence dashboard. We identify the need for role- and practice-sensitive support, experiment-level evidence structuring, inspectable and source-linked outputs, clear communication of uncertainty and coverage, and human oversight of quality and relevance judgments.
\end{itemize}

Together, our results provide insights for the development of future interactive systems that promote more efficient, transparent, and evidence-based research practices while supporting efforts to improve translation and reduce unnecessary animal experimentation.

\section{Related Work}

\subsection{Evidence Synthesis in Animal Research}

Evidence synthesis is a cornerstone of evidence-based medicine and has long been established in clinical research \cite{cookSystematicReviewsSynthesis1997}. Over the past two decades, increasing attention has also been given to rigorous evidence synthesis in animal research, where systematic reviews can help assess the robustness of existing findings, inform experimental design and model selection, and support implementation of the 3Rs \cite{ineichenSystematicReviewMetaanalysis2024a,devriesUsefulnessSystematicReviews2014b}. However, systematic reviews in both clinical and animal research remain resource-intensive and can require months of work \cite{borahAnalysisTimeWorkers2017a,bugajskaHowLongDoes2025}.

To reduce this burden, researchers have developed tools that automate parts of the evidence-synthesis process, including literature screening and classification \cite{marshallTowardSystematicReviewAutomation2019,bahorDevelopmentUptakeOnline2021c}, extraction of key study characteristics \cite{wangPICOEntityExtraction2022a,zurrer2024steed,doneva2025preclinie}, and organization of evidence into structured resources \cite{Doneva2026}. In animal research, initiatives such as the Systematic Review Facility (SyRF) have additionally explored interactive evidence platforms and topic-specific dashboards developed together with domain stakeholders \cite{bahorDevelopmentUptakeOnline2021c}. More recently, researchers have argued for broader evidence infrastructures that connect findings across animal and human research \cite{bannach-brownBuildingSynthesisreadyResearch2025,hairConnectingDotsNeuroscience2025a,riazFutureEvidenceSynthesis2024a}.

Yet comparatively little work has examined how evidence-support systems fit the broader practices and decision contexts of stakeholders in animal and translational research. Prior studies have evaluated the features, usability, and user experience of systematic-review tools, including tools used for screening and data extraction \cite{cleoUsabilityAcceptabilityFour2019,SoftwareToolsSystematic2024}. A recent multi-stakeholder workshop on evaluating preclinical models further identified barriers including workload, insufficient training, limited standardization, and poor integration with existing practices \cite{pistollatoEvaluatingTranslationalValue2026}. However, we still have limited empirical understanding of how different stakeholders work with evidence, where their needs diverge, and what they require from AI-assisted tools for translational evidence assessment.

Our work extends this line of research through in-depth interviews with stakeholders spanning basic and translational research, animal welfare and the 3Rs, and regulatory and ethical oversight. Rather than focusing on a single review task or disease-specific tool, we examine evidence practices and requirements across roles involved in animal-to-human translation. We additionally use a curated dashboard integrating animal and clinical evidence across neurological and psychiatric disease areas as a concrete probe to elicit requirements for more holistic evidence-support systems.

\subsection{AI-Assisted Scientific Literature Analysis}

Recent advances in AI and natural language processing have expanded the range of support available for scientific literature work, including semantic retrieval, relevance screening, and structured information extraction \cite{eger2025transforming,mikriukov2025ai,van2021open,peeters2025evaluation}. More recent LLM-based systems extend this support toward multi-step literature exploration, citation-backed synthesis, and scientific question answering \cite{skarlinski2024language,asai2026synthesizing,gao2025democratizing}. Prior work has also explored more specific research activities, such as retrieving potentially serendipitous scientific claims for hypothesis generation \cite{duckFindingNeedlesDocument2025b}.

Alongside these technical advances, HCI research has increasingly examined how AI-assisted literature tools fit researchers' workflows and expertise. Prior work has characterized practices and challenges around citation foraging, management, and synthesis \cite{fangExploringPracticesChallenges2025}, examined how LLMs support academic literature understanding and review \cite{wangEvaluatingLargeLanguage2024,choeSupportingNoviceResearchers2024}, and used user-centered design to develop tools for living reviews and interdisciplinary scoping reviews \cite{fok2025toward,mozgaiAcceleratingScopingReviews2024}. Together, this work shows that effective literature-support systems depend not only on model capabilities, but also on how they fit users' goals, expertise, and existing research practices.

Our work extends this literature to animal and translational research, where stakeholders across scientific, ethical, welfare, and regulatory roles engage with evidence for different purposes and decisions. We use a curated evidence dashboard as a concrete design probe to elicit how stakeholders would like evidence to be integrated, presented, and supported by automation. These findings can inform the design of more capable and context-sensitive AI-assisted evidence systems for translational research.

\subsection{Trust, Verification, and Human Oversight in AI-Assisted Evidence Work}

HCI research has shown that trust and effective use of AI depend on system, user, and contextual factors, including whether systems communicate their capabilities and limitations and enable users to inspect and question their outputs \cite{bach2024systematic,nushi2019guidelines,liao2020questioning}. Source-linked interfaces can support verification by connecting AI-generated content to its underlying evidence \cite{kambhamettu2025traceable}, while appropriate reliance requires users to accept useful AI support without over-relying on incorrect or uncertain outputs \cite{schemmer2023appropriate}.

These concerns are particularly important in evidence-intensive domains. Research on healthcare AI has shown that useful AI support must fit professional workflows, time constraints, domain expertise, and collaborative practices \cite{jacobs2021designing}. Work on generative AI for clinical evidence synthesis similarly identifies trustworthiness and verification as central concerns when producing automated evidence summaries \cite{zhang2024leveraging}. However, these questions remain comparatively underexplored in animal and translational research, where stakeholders must interpret heterogeneous and incomplete evidence and make context-dependent judgments about study quality, relevance, and translational value.

\medskip
Motivated by this gap, we examine how stakeholders in animal and translational research work with evidence and evaluate AI-assisted evidence support. We identify role- and decision-context-dependent practices, evidence-related challenges, and design and oversight requirements for future systems in this domain.

\section{Methods}
\label{sec:methods}

To explore current evidence practices, challenges, and design requirements for AI-assisted literature analysis in animal and translational research, we conducted semi-structured interviews with stakeholders from diverse professional backgrounds.

\begin{table*}[t]
\centering
\caption{Stakeholder categories used in the analysis for the 13 participants. Although participants could have experience across multiple categories, each was assigned to the category that best reflected their current professional role. 3Rs: Replacement, Reduction, and Refinement; GLP: Good Laboratory Practice. }
\label{tab:stakeholder-categories}
\small
\begin{tabular}{p{0.28\linewidth} p{0.08\linewidth} p{0.54\linewidth}}
\toprule
\textbf{Stakeholder category} & \textbf{N} & \textbf{Description} \\
\midrule
Basic research 
& 2
& Researchers conducting animal studies focused on investigating fundamental biological mechanisms and generating new scientific knowledge. \\

Translational research 
& 5
& Researchers using animal models to evaluate disease mechanisms and therapeutic approaches with the aim of informing clinical applications. \\

Animal welfare and 3R stakeholders 
& 4
& Participants focused on animal welfare, replacement, reduction, refinement, or the justification of animal studies. \\

Regulatory and ethics stakeholders 
& 2
& Participants involved in regulatory review, GLP, animal ethics, approval processes, or related oversight activities. \\
\bottomrule
\end{tabular}
\end{table*}

\subsection{Participants}

We recruited 13 participants for this study. 
Participants were identified through professional networks and purposive sampling to capture a range of perspectives across animal and translational research. We included participants if they had experience using scientific literature to inform research, evaluation, oversight, or decision-making related to animal or translational research.

Based on their primary professional perspective, we grouped participants into four categories: basic research, translational research, animal welfare and the 3Rs, and regulatory and ethics roles. Table~\ref{tab:stakeholder-categories} defines these categories.

Recruitment aimed to capture perspectives across the main stakeholder groups relevant to our research questions. The resulting interviews provided a sufficiently rich dataset to examine variation in evidence practices, challenges, and requirements across these groups. Table~\ref{tab:participants} summarizes the 13 interview participants included in the analysis. 

All participants were based in Switzerland at the time of the interview. Of the 13 participants, five identified as women and eight as men.

\begin{table*}[t]
\centering
\caption{Overview of interview participants. Participants are anonymized and grouped by their primary stakeholder perspective.}
\label{tab:participants}
\small
\begin{tabular}{p{0.02\linewidth} p{0.18\linewidth} p{0.32\linewidth} p{0.40\linewidth}}
\toprule
\textbf{ID} &
\textbf{Stakeholder perspective} &
\textbf{Current role} &
\textbf{Primary expertise} \\
\midrule
P01
& Translational research
& Postdoctoral researcher
& Preclinical disease models and therapeutic evaluation \\
P02 
& Animal welfare and 3Rs
& University animal welfare officer
& Animal welfare oversight and experimental neuroscience \\
P03 
& Regulatory and ethics
& Preclinical reviewer and GLP inspector
& Nonclinical safety assessment and GLP compliance \\
P04 
& Animal welfare and 3Rs
& Veterinarian at an animal protection organization
& Animal welfare policy and advocacy \\
P05 
& Basic research
& Research staff member
& Experimental neuroscience and animal behaviour \\
P06 
& Basic research
& Professor
& Neurogenesis and neuroscience research leadership \\
P07 
& Translational research
& Professor
& Behavioural neurobiology and psychiatric disease models \\
P08 
& Translational research
& Doctoral researcher
& Preclinical oncology and animal model refinement \\
P09 
& Animal welfare and 3Rs
& Scientific officer and data analyst
& 3Rs research and animal-use data analysis \\
P10 
& Animal welfare and 3Rs
& Corporate animal welfare officer
& Corporate animal welfare and 3Rs implementation \\
P11 
& Regulatory and ethics
& Scientific collaborator at a federal authority
& Regulatory review of animal experiment applications \\
P12 
& Translational research
& Biomedical data scientist
& Biomedical data science and literature mining \\
P13 
& Translational research
& Pharmaceutical research scientist
& Preclinical drug development and translational research \\
\bottomrule
\end{tabular}
\end{table*}

\subsection{Procedure and Analysis}

\paragraph{Interviews.}
Interviews were conducted and recorded via Zoom after participants provided informed consent. We first introduced the study and asked brief questions about participants' backgrounds before following the interview protocol (see Appendix \ref{app:interview-protocol}). Each interview lasted around one hour.

The semi-structured interviews consisted of two stages: The first examined participants' literature workflows, information needs, and challenges, including how they searched for evidence, evaluated and compared studies, connected animal and human evidence, and used systematic reviews, literature databases, and AI-assisted tools. 

In the second stage, participants explored an interactive dashboard for animal and translational research evidence while thinking aloud. We then asked about its usefulness, missing functionality, trustworthiness, and potential use cases in supporting translational research and the 3Rs.

\paragraph{Analysis.}
We analyzed the automatically transcribed interview transcripts using reflexive thematic analysis \cite{braunUsingThematicAnalysis2006,braunReflectingReflexiveThematic2019}, supported by the qualitative data analysis software QualCoder \cite{curtain2025qualcoder}. Two researchers (co-authors of this paper) first familiarized themselves with the data by independently reading and coding the same two transcripts. Coding was primarily inductive and focused on segments relevant to the research questions. The researchers then compared their interpretations and discussed similarities and differences in their initial codes to develop a shared conceptual understanding of the data. Each researcher subsequently coded approximately half of the remaining transcripts, while remaining open to refining and extending the initial codes.

Theme development was conducted as collective work in the author team. The researchers brought the codes together on a Miro board \cite{miro2026} and, over several meetings, iteratively organized and re-organized them into candidate themes based on recurring patterns of meaning across participants. Candidate themes were subsequently reviewed in relation to the research questions and the underlying coded excerpts. For each theme, the researchers returned to the associated interview excerpts, collated illustrative quotations, and jointly developed the interpretation presented in the findings. Examples of the transcript coding, evolving codebook, and collaborative theme development are provided in Appendix~\ref{app:qualitative-analysis}.

\paragraph{Positionality.}
We acknowledge that our backgrounds and institutional context shaped how we conceptualized the study, recruited participants, and interpreted the data. Our team’s experience in medical data science, AI, and animal and translational research informed the interview topics, particularly evidence assessment, translational relevance, and opportunities for AI support. To limit the influence of these prior assumptions, we began with open-ended questions about participants’ practices and challenges before asking about specific technologies or design possibilities. Participants were recruited primarily through academic and professional networks in Switzerland, supplemented by snowball sampling. Consequently, our findings might reflect the practices and regulatory context of animal and translational research in Switzerland.

\subsection{Curated Evidence Dashboard}

As a concrete design probe, participants explored an interactive dashboard developed as a Shiny app. The dashboard provided access to a curated dataset linking preclinical neuroscience publications with clinical trials through drug--disease relationships. The underlying data were generated through an NLP-based pipeline described in work by~\citet{donevaLargeScaleAssessmentAnimaltoHuman2026}, which integrates animal studies from PubMed with ClinicalTrials.gov and regulatory data.

\begin{figure*}[t]
    \centering
    \includegraphics[width=\textwidth]{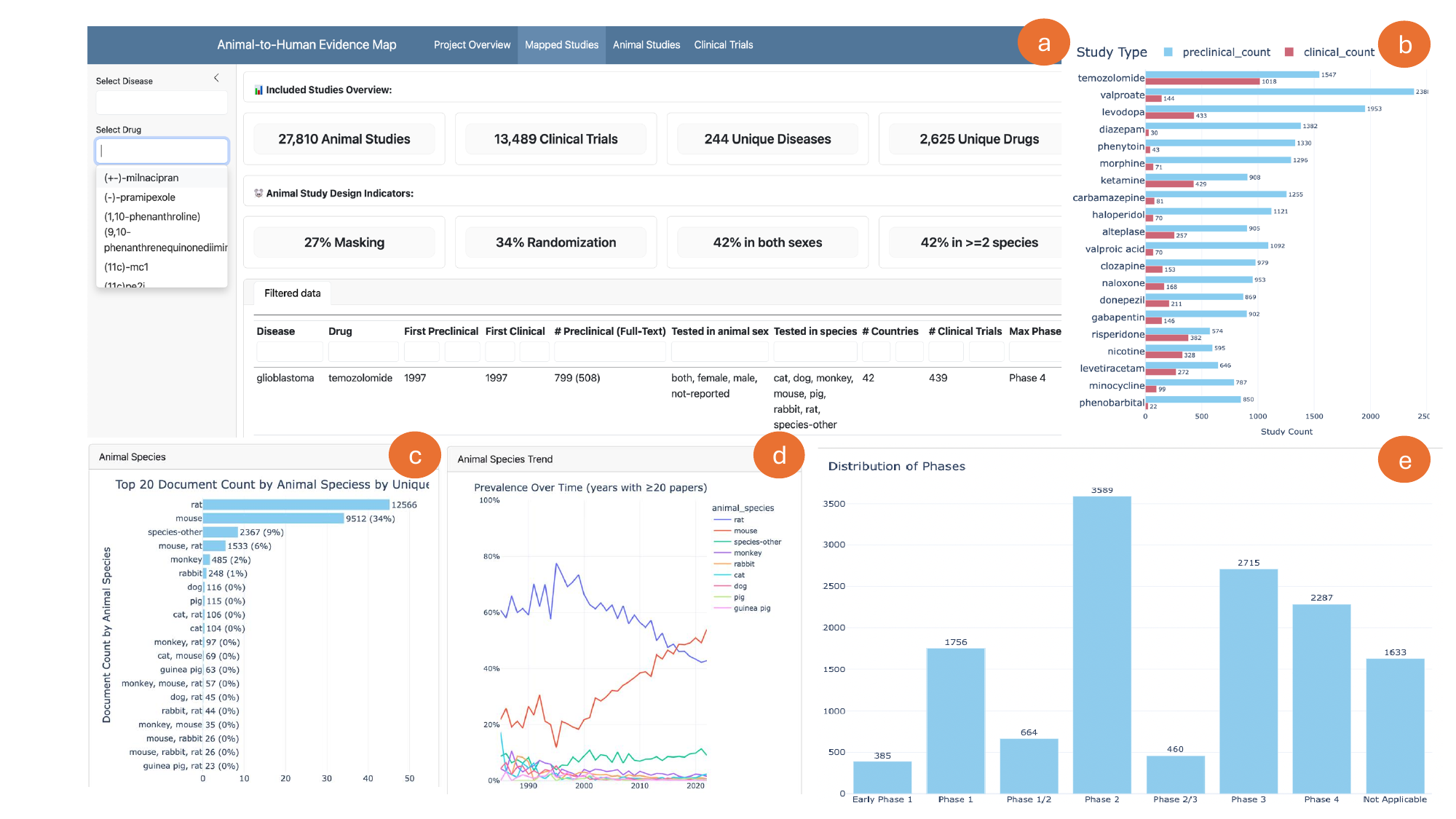}
    \caption{
    Overview of the \textit{Animal-to-Human Evidence Map} dashboard used during the interviews.
    (a) The main dashboard provides disease- and drug-level filtering, summary indicators of the available preclinical and clinical evidence, animal study design indicators, and a study-level comparison table.
    (b) Comparison of preclinical and clinical study counts across frequently studied drugs.
    (c) Distribution of animal species represented in the preclinical evidence.
    (d) Changes in the prevalence of reported animal species over time.
    (e) Distribution of clinical trials across development phases.
    }
    \label{fig:dashboard}
    \Description{Five dashboard views summarize available animal and clinical evidence and reporting patterns. The main view combines disease and drug filters with study counts, reporting indicators, and a study-level comparison table. Supporting charts compare preclinical and clinical study volumes across drugs, show species composition and changes over time, and summarize clinical trials by development phase.}
\end{figure*}

Users could filter the evidence by disease and drug, inspect linked animal and clinical studies, and explore study counts, timelines, translational patterns, and selected characteristics of the underlying animal experiments. These included species, strain, sex, animal numbers, assay types, and the reporting of welfare, masking, randomization, and sample-size calculations. Figure~\ref{fig:dashboard} provides an overview of the interface and examples of its main visualizations.

\section{Results}
\subsection{Stakeholder Practices and Challenges in Evidence Work}

\begin{table*}[t]
\centering

\begin{threeparttable}

\caption{Evidence-work challenges across stakeholder interviews.
Filled circles indicate interviews that contributed evidence for a
challenge; hollow circles indicate that the challenge was not identified.}

\label{tab:challenges-circles}

\footnotesize
\setlength{\tabcolsep}{2.5pt}
\renewcommand{\arraystretch}{1.45}

\begin{tabularx}{\textwidth}{
  @{}
  >{\raggedright\arraybackslash}X
  >{\centering\arraybackslash}p{0.095\textwidth}
  >{\centering\arraybackslash}p{0.185\textwidth}
  >{\centering\arraybackslash}p{0.155\textwidth}
  >{\centering\arraybackslash}p{0.105\textwidth}
  @{}
}

\toprule

\textbf{Challenge}
&
\cellcolor{basic!18}
\shortstack[c]{
  \textcolor{basic!85!black}{\textbf{Basic research}}\\[-1pt]
  \scriptsize P05 \quad P06
}
&
\cellcolor{trans!18}
\shortstack[c]{
  \textcolor{trans!85!black}{\textbf{Translational research}}\\[-1pt]
  \scriptsize P01 \quad P07 \quad P08 \quad P12 \quad P13
}
&
\cellcolor{threeR!20}
\shortstack[c]{
  \textcolor{threeR!85!black}{\textbf{Animal welfare}}\\[-1pt]
  \textcolor{threeR!85!black}{\textbf{\& 3Rs}}\\[-1pt]
  \scriptsize P02 \quad P04 \quad P09 \quad P10
}
&
\cellcolor{reg!18}
\shortstack[c]{
  \textcolor{reg!85!black}{\textbf{Regulatory}}\\[-1pt]
  \textcolor{reg!85!black}{\textbf{\& ethics}}\\[-1pt]
  \scriptsize P03 \quad P11
}
\\

\midrule

\stage{Find and access evidence}
\hspace{5pt}Specialist literature not indexed in standard databases
& \Bcell{\oB}{\oB}
& \Tcell{\oT}{\oT}{\oT}{\oT}{\oT}
& \Rcell{\dR}{\oR}{\oR}{\dR}
& \Ecell{\oE}{\oE}
\\
\hspace{5pt}Systematic search impractical at scale
& \Bcell{\dB}{\oB}
& \Tcell{\oT}{\oT}{\dT}{\dT}{\oT}
& \Rcell{\dR}{\oR}{\dR}{\oR}
& \Ecell{\oE}{\oE}
\\
\hspace{5pt}Negative results absent from the published record
& \Bcell{\dB}{\dB}
& \Tcell{\dT}{\oT}{\dT}{\oT}{\oT}
& \Rcell{\dR}{\oR}{\oR}{\oR}
& \Ecell{\oE}{\oE}
\\
\hspace{5pt}Evidence inaccessible because of commercial, regulatory,
or institutional constraints
& \Bcell{\oB}{\oB}
& \Tcell{\oT}{\oT}{\oT}{\oT}{\dT}
& \Rcell{\oR}{\oR}{\oR}{\oR}
& \Ecell{\dE}{\dE}
\\[2pt]

\stage{Appraise evidence}
\hspace{5pt}Appraisal criteria not transferable across stakeholder roles
& \Bcell{\dB}{\dB}
& \Tcell{\dT}{\dT}{\dT}{\dT}{\dT}
& \Rcell{\dR}{\dR}{\oR}{\dR}
& \Ecell{\dE}{\dE}
\\
\hspace{5pt}Expert judgment required even with structured frameworks
& \Bcell{\dB}{\oB}
& \Tcell{\oT}{\dT}{\oT}{\dT}{\dT} 
& \Rcell{\oR}{\oR}{\oR}{\oR}
& \Ecell{\oE}{\oE}
\\[2pt]

\stage{Compare studies}
\hspace{5pt}Experimental parameters
(species, dose, sample size) missing
& \Bcell{\dB}{\oB} 
& \Tcell{\dT}{\dT}{\dT}{\oT}{\dT} 
& \Rcell{\dR}{\oR}{\dR}{\dR}
& \Ecell{\oE}{\oE}
\\
\hspace{5pt}Drug formulation and administration procedures unclear
& \Bcell{\oB}{\dB}
& \Tcell{\dT}{\oT}{\oT}{\oT}{\oT}
& \Rcell{\oR}{\oR}{\oR}{\oR}
& \Ecell{\oE}{\oE}
\\
\hspace{5pt}Conditions too variable for cross-laboratory comparison
& \Bcell{\dB}{\dB}
& \Tcell{\oT}{\oT}{\oT}{\oT}{\dT}
& \Rcell{\oR}{\oR}{\oR}{\oR}
& \Ecell{\oE}{\oE}
\\[2pt]

\stage{Assess welfare and 3R evidence}
\hspace{5pt}Welfare parameters
(housing, anaesthesia) inconsistently reported
& \Bcell{\oB}{\oB}
& \Tcell{\oT}{\oT}{\dT}{\oT}{\dT} 
& \Rcell{\dR}{\oR}{\oR}{\dR}
& \Ecell{\oE}{\oE}
\\
\hspace{5pt}Cannot determine whether refinement was tested or unreported
& \Bcell{\oB}{\oB}
& \Tcell{\oT}{\oT}{\dT}{\oT}{\oT}
& \Rcell{\dR}{\oR}{\oR}{\oR}
& \Ecell{\oE}{\dE}
\\[2pt]

\stage{Synthesise across species and domains}
\hspace{5pt}Cross-species biological and outcome correspondence uncertain
& \Bcell{\dB}{\oB}
& \Tcell{\oT}{\dT}{\dT}{\oT}{\dT}
& \Rcell{\oR}{\oR}{\oR}{\oR}
& \Ecell{\dE}{\dE}
\\[2pt]

\stage{Use AI tools}
\hspace{5pt}AI generates fabricated citations or misrepresents sources
& \Bcell{\dB}{\oB}
& \Tcell{\dT}{\oT}{\oT}{\oT}{\dT} 
& \Rcell{\oR}{\oR}{\dR}{\dR}
& \Ecell{\oE}{\oE}
\\
\hspace{5pt}Generated claims lack source-level traceability
& \Bcell{\dB}{\oB} 
& \Tcell{\oT}{\dT}{\oT}{\dT}{\dT} 
& \Rcell{\oR}{\oR}{\oR}{\oR}
& \Ecell{\oE}{\oE}
\\
\hspace{5pt}AI risks systematic overstatement of translational evidence
& \Bcell{\oB}{\oB}
& \Tcell{\oT}{\oT}{\oT}{\oT}{\dT}
& \Rcell{\oR}{\oR}{\oR}{\oR}
& \Ecell{\oE}{\oE}
\\

\bottomrule
\end{tabularx}

\begin{tablenotes}[para]
\small
\item Circle positions correspond to the participant IDs shown in each
stakeholder-group header. Filled circles indicate interviews contributing
evidence for the challenge. Hollow circles indicate only that the challenge
was not identified in that interview; they do not indicate disagreement or
absence of the challenge in practice.
\end{tablenotes}

\end{threeparttable}
\end{table*}

\subsubsection{\textbf{Evidence Use and Appraisal Depended on Role and Decision Context}}

Participants used literature differently depending on their professional role and the task at hand. Researchers drew on published evidence to plan experiments, check whether similar work had already been done, and select suitable models or procedures \textit{(P01, P05, P07, P08, P13)}. In supervisory and regulatory work, literature engagement meant reviewing evidence assembled by junior researchers or submitted by applicants rather than conducting primary searches \textit{(P03, P05, P11)}. Participants working in animal welfare and 3R contexts focused on monitoring procedures, welfare scoring systems, and refinement methods \textit{(P02, P04, P10, P11)}.

Participants also applied relevance judgments before examining papers in detail. P07 assessed model suitability before reading a paper in detail, excluding studies using tests they considered irrelevant to the human condition being modelled:

  \begin{quote}
  ``If a paper is based on tests which I have the opinion are not relevant,
  then I won't read it.'' \textit{(P07)}
  \end{quote}

Once papers had been selected, extraction followed the same role-specific logic: experimental parameters for researchers, welfare and refinement information for 3R participants, and structured compliance and risk-of-bias data for regulatory and systematic-review work \textit{(P01--P04, P07, P08, P10--P13)}. The methods section was frequently valued because it revealed ``the meat of what people have actually done'' \textit{(P09)}.

Evidence appraisal relied on criteria grounded in participants' disciplines and decision contexts rather than on a shared quality standard. Regulatory appraisal emphasized Good Laboratory Practice (GLP) compliance, independent replication, and detailed study reports \textit{(P03, P11)}, whereas translational and pharmaceutical appraisal considered biological responses, drug effects, and model relevance \textit{(P07, P13)}. Because these criteria were not directly transferable across roles, the same study could be judged differently depending on who assessed it and for what purpose.

Even within similar context, assessing evidence quality required case-by-case judgment. P05 described methodological limitations that became visible only during replication work and were not apparent in a high-impact publication. Structured assessment frameworks provided guidance but did not remove the need for expert judgment. P12 used the Cochrane Risk of Bias 2 (RoB~2) framework, which helps reviewers assess whether features of a study's design, conduct, or reporting could systematically distort its findings \cite{minozzi2020revised}. Despite using this structured framework, trained assessors still disagreed about how to rate individual studies:

\begin{quote}
  ``Scientists did not agree. They used two to three hours to debate on whether a study was high risk or moderate risk.'' \textit{(P12)}
\end{quote}

Evidence appraisal therefore remained dependent on role-specific expertise and case-by-case judgment, even when participants used structured frameworks \textit{(P05, P07, P12, P13)}.

\subsubsection{\textbf{Participants Relied on Selective Search and Personal Evidence-Management Strategies}}
Because the volume of literature exceeded what participants could feasibly review, fully systematic searches were considered too costly for many research decisions \textit{(P02, P05, P08, P09, P12)}.

Therefore, participants searched selectively rather than comprehensively. They reported relying mainly on familiar search platforms, particularly PubMed and Google Scholar, supplemented by citation chains, professional networks, conference materials, and monitoring of specialist journals or research groups \textit{(P01--P09, P12, P13)}. Some participants prioritised papers using citation counts, journal reputation, study recency, or laboratory track record
\textit{(P01, P03--P06, P09)}, before screening abstracts and examining relevant papers more closely.

To manage and compare this evidence, participants developed personal systems rather than adopting shared tools. These included Excel-based scoring schemes \textit{(P01, P02)}, a customised reference-manager taxonomy organised by species, brain region, and behaviour \textit{(P07)}, and a computational pipeline covering retrieval, screening, and
risk-of-bias appraisal \textit{(P12)}. 

Overall, participants adapted to literature overload by combining selective search heuristics with personal evidence-management systems, resulting in heterogeneous and largely individualized review practices.

\subsubsection{\textbf{Evidence Was Absent, Inaccessible, or Incompletely Reported}}
Participants described fundamental limitations in the availability and reporting of evidence. Studies reporting null, negative, or unsuccessful outcomes were perceived to be systematically underrepresented in the published record \textit{(P01, P02, P05, P08)}. P05 framed the distinction explicitly:

  \begin{quote}
  ``The hardest thing to find are the negative results. But I don't think
  they're hard to find; I think they're just not present.'' \textit{(P05)}
  \end{quote}

Information about unsuccessful experiments or development efforts sometimes circulated informally through professional networks \textit{(P06)}. Other relevant evidence remained inaccessible because of commercial, regulatory, or institutional constraints \textit{(P03, P11, P13)}. P11, for example, lost access to databases and full-text publications after moving between organisations. 

Even when relevant studies could be found, key study parameters were often poorly reported or missing, limiting evaluation and comparison across studies. Basic experimental details, including species, strain, sex, dose, and sample size, were frequently missing or inadequately documented in the methods \textit{(P01, P02, P05, P07-P10, P13)}. Drug formulation, additives, and administration procedures were similarly unclear \textit{(P01, P06)}.

Welfare and 3R-related information, including body weight, housing, anaesthesia, and refinement procedures, was reported inconsistently, and often in supplementary materials only \textit{(P02, P08, P10, P13)}. For some animal welfare and 3R participants, missing welfare and refinement details complicated licence preparation and refinement decisions. Participants could not always determine whether a refinement had been tested and found unsuitable or had simply gone unreported
\textit{(P02, P08, P11)}.

\subsubsection{\textbf{Cross-Study and Cross-Species Synthesis Required Interpretive Judgment}}

For participants in translational or 3R contexts, synthesis required connecting evidence across domains. Translational participants looked for correspondence between biological mechanisms, model characteristics,
experimental outcomes, and human conditions rather than identical experiments \textit{(P05, P07, P11, P13)}. Participants working in 3R contexts combined experimental findings with welfare and refinement evidence to assess whether less burdensome alternatives were available \textit{(P02, P04, P10, P11)}. Establishing these connections could also require collaboration across disciplines; P06 described human and rodent researchers working together to define comparable behavioural endpoints.

Comparison remained difficult even when studies addressed the same disease, intervention, or broad animal model. Differences in animal characteristics, experimental conditions, procedures, outcome measures, and analysis meant that apparently similar studies could not always be treated as equivalent. P05 described the extent of this variation in neuroscience:

\begin{quote}
``The task is different, the place is different, the holding facilities are
different, the behaviour is different, the way you implant, the way you
analyse, everything is difficult to compare.'' \textit{(P05)}
\end{quote}

Cross-species comparison introduced further uncertainty. Physiological measures could be relatively comparable, whereas cognitive and behavioural functions were harder to connect across species \textit{(P03, P05, P07)}. Animal and human studies could also use different measures of success, such as tumour regression in animal experiments and five-year survival in patients \textit{(P08)}. Participants further lacked quantitative evidence needed for translational appraisal, including effect sizes, dose--response relationships, and data on how reliably particular models translated to human outcomes
\textit{(P13)}.

Translational synthesis therefore depended on interpreting which similarities and differences across heterogeneous studies and species were meaningful for the translational question at hand.

\subsubsection{\textbf{AI Use Varied, but Reliability Concerns Limited Trust}}

Participants incorporated AI into literature work to varying degrees. P07 rejected AI-assisted extraction because it conflicted with his close-reading practice, whereas P13 had largely replaced manual searching with AI-supported workflows using Gemini Deep Research~\cite{comanici2025gemini} and EMET from BenchSci~\cite{benchsci2026emet}.

AI was most commonly used to generate search strings or obtain initial topic overviews \textit{(P01, P05, P08, P11, P13)}. P11 described how LLMs made PubMed query formulation much faster, while reading the retrieved papers remained the main time-consuming task:

\begin{quote}
  ``Before LLMs was the search on PubMed, and now that I can have strings
  written in a couple of minutes, seconds. Makes things much faster. I think
  the most time-consuming thing is reading the stuff that I find.''
  \textit{(P11)}
\end{quote}

Participants also used AI for abstract screening, paper summarisation, documentation, and querying uploaded collections of papers or licence documents \textit{(P04, P06, P08--P10, P13)}. P08 found that, in a systematic screening of several thousand papers, AI performed worse than expected on simple yes--no inclusion decisions but better on more complex ones.

These benefits were accompanied by reliability concerns. Participants encountered fabricated publications, incorrect citations, and inaccurate summaries in AI-generated outputs, shifting effort from discovery toward verification. They therefore checked outputs against original publications or other tools \textit{(P01, P05, P09, P10, P13)}, and P10 stopped using AI for literature discovery altogether after encountering too many hallucinated sources. P05 described the central check:

\begin{quote}
``The important part is to really go and check if the publications that are
given by the LLM are actually saying what the LLM is saying.''
\textit{(P05)}
\end{quote}

Participants also questioned whether AI-assisted searches returned complete or unbiased results: P09 reported never relying on a single AI tool alone out of concern that it would miss relevant studies.

Some risks could not be addressed through source checking alone. P13 raised the possibility that AI might systematically frame evidence as more translatable than the underlying studies supported---a directional bias
requiring domain expertise rather than simple cross-referencing to detect. Limited source-level traceability further complicated verification: without links from generated claims to source passages, participants found it difficult to assess whether summaries were accurate \textit{(P05, P07, P12, P13)}.

AI accelerated parts of evidence work but shifted effort toward verification and expert interpretation. These findings informed the design requirements discussed next.

\subsection{Information Needs and Design Requirements}
Participants described both specific information needs and broader requirements for AI-assisted evidence systems. Table~\ref{tab:needs-design-implications} summarizes the resulting design implications, which are elaborated below.

\subsubsection{\textbf{Support Tools Should Adapt to the User’s Research Context}}

Several participants wanted tools to account for the user's role, research question, 
and experimental context (P05, P06, P10). 

To illustrate this, one participant envisioned a context-sensitive ranking based on relevance:

\begin{quote}
    ``Running an LLM in the background, where I can say, `Hey, look, I am this 
    kind of person, I do this kind of research. Could you please grade the 
    studies based on uncertainty and on how well they align with my research?' 
    Then I could get a prompt saying, `This could be relevant for you because 
    of this and this reason.' Ideally, it would also tell me why it is relevant 
    for me.''
    \hfill\textit{[P05]}
\end{quote}

Participants also envisioned an interactive process in which the system first 
provided an overview and then supported follow-up questions. For example, P06 
wanted to examine dominant approaches in a field, understand why researchers 
used different methods, and explore why apparently similar studies reached 
different conclusions. 

Personalization also extended beyond individual queries.
P10 envisioned a persistent, user-specific workspace where he could ``run searches, prompt the system, keep papers, and cross-reference them'' within a single topic area.

\begin{table*}[t]
\caption{Participant needs and corresponding design implications for future AI-assisted biomedical evidence systems.}
\label{tab:needs-design-implications}

\footnotesize
\setlength{\tabcolsep}{5pt}
\renewcommand{\arraystretch}{1.15}

\begin{tabularx}{\textwidth}{
@{}
L{0.20\textwidth}
L{0.25\textwidth}
Y
@{}
}
\toprule
\textbf{Design area} &
\textbf{Participant need} &
\textbf{Design implications} \\
\midrule

\rowcolor{gray!12}
\multicolumn{3}{@{}l}{\textbf{Context-sensitive interaction}} \\

Personalized and persistent support &
Evidence adapted to the user's role, research question, and project context &
Adapt retrieval and ranking to the user's context, explain why studies are relevant, support follow-up questions, and provide persistent workspaces for storing and cross-referencing evidence across sessions. \\

\midrule

\rowcolor{gray!12}
\multicolumn{3}{@{}l}{\textbf{Experiment-level evidence}} \\

Structured experimental representation &
Direct access to the individual experiments reported within publications &
Represent distinct experimental arms separately and extract key information on models, procedures, interventions, outcomes, quantitative results, and welfare conditions. \\

\midrule

\rowcolor{gray!12}
\multicolumn{3}{@{}l}{\textbf{Cross-domain evidence exploration}} \\

Semantic and relational support &
Access to related evidence beyond exact keyword matches &
Normalize terminology and connect related models, mechanisms, interventions, readouts, and clinical outcomes across animal and human research. Support comparison of conflicting, negative, and translational findings. \\

\midrule

\rowcolor{gray!12}
\multicolumn{3}{@{}l}{\textbf{Trustworthy assessment}} \\

Inspectable evidence and human judgment &
Reliable support without replacing contextual scientific interpretation &
Link extracted and generated information to its supporting evidence, disclose coverage and uncertainty, distinguish missing reporting from failed extraction, and preserve user control over quality and relevance judgments. \\

\midrule

\rowcolor{gray!12}
\multicolumn{3}{@{}l}{\textbf{Traceable outputs}} \\

Evidence-grounded synthesis &
Faster production of reviews, reports, and licence-related text &
Generate outputs from user-selected evidence collections while preserving links between each claim and the underlying publication, experiment, and source passage. \\

\bottomrule
\end{tabularx}
\end{table*}

\subsubsection{\textbf{Support Tools Should Structure Evidence at the Experiment Level}}
\label{sec:experiment-level-evidence}

Participants wanted evidence represented at the level of individual
experiments rather than only at the publication level (P13, P05, P01). P01 described how experiment-level structuring could be helpful:

\begin{quote}
``Something that helps to isolate the animal studies in each paper without
me going through the full paper would be very helpful. Sometimes the paper
is messy and the experiment is not clearly presented. If it could give you
the output of what the mouse experiment was about, with all the details and
numbers, this would speed up the workflow so much.''
\hfill\textit{[P01]}
\end{quote}

The requested information included general publication metadata, as well as key experimental characteristics. Where a publication reported multiple experimental arms, these should be represented separately rather than aggregated at the publication level. A complete overview of the expected fields mentioned by all participants is provided in Table~\ref{tab:desired-extracted-characteristics} in the Appendix.

\subsubsection{\textbf{Support Tools Should Connect and Evaluate Evidence Across Research Domains}}
\label{sec:semantic-relational-exploration}

Participants wanted support tools to move beyond exact keyword matching by recognizing equivalent or related terminology across fields (P09, P02, P04, P12). This was particularly important when linking animal and human evidence, where the same concepts may be described differently. As P09 noted, ``a first step is always to check that you have the right terms, or the right understanding of how people call the thing.''

Beyond terminology normalization, participants wanted tools to reveal relationships among animal models, targets, mechanisms, compounds, readouts, and clinical outcomes (P03, P11, P13). They also wanted to identify contradictory findings and understand why apparently similar studies produced different results (P05, P06, P07, P12). P09 illustrated how a system might surface a relationship not specified in the initial query:

\begin{quote}
``A neurotransmitter has been studied in the context of this disease, and
then it could say, `It is actually often studied with this other mechanism
or this other neurotransmitter, so you should probably look at both
together...' ''
\hfill\textit{[P09]}
\end{quote}

These connections were particularly relevant for translational research. Participants described a need to trace clinical observations back to the underlying animal experiments and examine whether related findings had already been reported but overlooked (P02, P10, P01):

\begin{quote}
``There is no unified system where a clinician could go back to the animal
studies that were conducted and read how they were conducted...''
\hfill\textit{[P10]}
\end{quote}

Participants also wanted these connections to inform decisions about proposed animal experiments. They saw value in comparing new proposals with similar prior studies to assess their justification and identify possible improvements (P05, P02). P02 envisioned a system that could compare an application for approval of an animal experiment with the existing literature and indicate whether ``this was already done,'' ``this refinement is possible,'' or ``a replacement has already been established.''

Participants differed in how far systems should move from identifying relationships to interpreting them. Some viewed semantic linking and structured comparison as the most realistic contribution, while others envisioned support for evaluating translational success, explaining conflicting results, or considering how negative findings should inform future work (P06, P05).

\subsubsection{\textbf{Trustworthy Evidence Support Requires Human Judgment and Inspectable Outputs}}

Participants did not expect support systems to replace scientific interpretation. They distinguished between extracting how an experiment was conducted and interpreting why particular models, methods, or hypotheses were chosen. As P01 noted, ``understanding why they did it is more of a theoretical job.''

Participants were particularly cautious about automated judgments of study quality (P02, P05, P06). Such assessments depended on the research question and experimental context. P05 therefore emphasized the need for human oversight:

\begin{quote}
``How does a tool decide what is good and bad research? Maybe it is good
research for one person, but bad research for another. I think the human
still needs to be in the loop...''
\hfill\textit{[P05]}
\end{quote}

At the same time, participants envisioned systems generating outputs such as systematic reviews or text for animal licence applications. For such outputs to be useful, substantive claims needed to remain linked to the publications, experiments, or source passages supporting them \textit{(P05, P07, P12)}. Participants also expected systems not to invent papers, citations, or claims and to communicate the scope of the literature searched, missing evidence, and uncertainty \textit{(P04, P09, P12, P13)}.

The value of this support lay partly in speed. Rather than expecting a definitive answer, P09 wanted a system to assemble ``not actually the answer, but the literature'' within a few days of receiving a stakeholder question. Such tools could therefore accelerate the transition from evidence collection to analysis and communication, while leaving contextual assessment and final conclusions to human experts.

\subsection{Stakeholder Perceptions of the Dashboard}
\subsubsection{\textbf{Participants Perceived the Dashboard as Supporting Rapid Study Screening}}

Participants generally perceived the dashboard as clear, intuitive, and easy to navigate after a brief introduction (P01, P05, P09, P03, P07, P08, P11, P12). They particularly valued its ability to condense a large evidence base into an accessible overview (P01, P05, P04). As P01 explained:

\begin{quote}
``It really gives you a snapshot of all the studies that you have. It makes
your mind clear.''
\hfill\textit{(P01)}
\end{quote}

Participants reported that the tables, filters, and visualizations made broad reporting patterns and study characteristics easier to identify without first reading every publication in full (P01, P05, P09, P02). They viewed this information as useful for deciding which publications warranted closer inspection. P02 noted that the dashboard allowed them to ``judge the quality of the information more quickly and decide whether to look more deeply into [a] publication.''

\subsubsection{\textbf{The Dashboard Required More Detailed and User-Specific Exploration}}

Although participants generally valued the dashboard as an accessible overview, its usefulness depended on their role, prior knowledge, and decision context. 

The disease--drug structure aligned well with clinical and translational questions but was less suitable for basic research. P05 noted that basic researchers would need ``many more filters,'' and proposed keyword search across titles, abstracts, or full texts, to explore brain regions, behavioural phenotypes, cognitive concepts, and experimental methods.

Pharmaceutical and regulatory participants also saw potential applications, but emphasized that these would require greater detail (P03, P11, P12, P13). Participants therefore requested including more of the experiment-level characteristics discussed in Section~\ref{sec:experiment-level-evidence} (P05, P02, P09). As P13 highlighted:

\begin{quote}
``I would very much start with this tool, to be honest. But right now, I see
that it stops at the layer where it becomes interesting. I am missing one or
two additional layers of detail.''
\hfill\textit{(P13)}
\end{quote}

Furthermore, participants wanted greater control over which information was displayed and how studies were organized. P05 proposed personalized filters so that ``the information that comes out would be tailored to me,'' allowing the interface to adapt to different research questions and evidence needs. A detailed overview of the requested search, filtering, and prioritization functions is provided in Table~\ref{tab:desired-search-functions} in the Appendix.

\subsubsection{\textbf{Further Analysis Should Be Supported In-Platform or Through Interoperability}}

Participants wanted either sufficient functionality to organize and analyse evidence within the dashboard or flexible transfer into existing workflows (P01, P05, P02, P09, P11). Desired in-platform functions included sorting, filtering, notes taking and custom ratings. Further requested features included retained evidence collections and shared institutional workspaces through which research could be revisited and shared with supervisors, students, or collaborators (P01, P05, P02).

Where these functions were unavailable, participants requested export to spreadsheets, reference managers, machine-readable formats, and APIs (P02, P09).
They also suggested integration with institutional and external evidence resources, including links to the original publications, registries, omics repositories, and literature or regulatory databases (P05, P01, P03, P07, P12, P11, P13). 

\subsubsection{\textbf{Trust Depends on Auditability, Coverage, and Independent Testing}}
Participants reported a relatively high baseline trust in research tools (P01, P05), in particular with university-developed systems (P03). Nevertheless, they expected to verify the dashboard before relying on it (P11, P01). Initial checks involved comparing a small sample of dashboard outputs with the original publications:

\begin{quote}
``At the beginning, I would open a couple of studies---two, three, four
studies---and see whether the information is really the same. If the first
studies all matched, and things continued to work well, then I would say,
okay, this is trustworthy.''
\hfill\textit{(P01)}
\end{quote}

Source-level auditability was therefore important. Participants wanted extracted values and summaries to remain linked to their supporting passages so that visible errors could be checked and corrected (P05, P07, P12).

Retrieval coverage was harder to assess. As P01 explained, ``anything else is easy to check by clicking. But how well it finds the studies---that is the only point that is hard to know and hard to verify'' (P01, P04). Participants therefore expected to compare the dashboard with PubMed or other search systems and use it as a first-line or complementary source rather than rely on it alone (P01, P04, P03, P07, P08).

Trust also depended on transparency about coverage and maintenance. Participants requested clear information about the databases, publication types, date ranges, and latest update included in the system, together with date filters and continued updating (P02, P04, P05, P06).

\subsubsection{\textbf{Potential 3R Value Depends on Concrete Decision Contexts}}

Participants located the dashboard's potential 3R impact in more optimal animal experiment planning and the assessment of proposed animal studies. Faster access to previous experiments could help researchers learn from existing methods and outcomes. It could support better-informed study design and reduce avoidable weaknesses in new experiments (P01, P05, P09).

Participants also saw potential for more explicit 3R support. This would require extending the dashboard with structured information on animal welfare related aspects. Such additions could support licence preparation and help link alternatives to relevant disease models (P05, P02, P08, P13).

The dashboard could also support assessment of whether the expected scientific or societal benefits of an animal experiment justify its potential harms by linking animal studies with related human clinical trials and providing a view of a model's translational performance. Participants saw this as a useful starting point for assessing model justification, although more detailed evidence would be needed for formal review (P03, P11).

\section{Discussion}

Across interviews, evidence use in animal and translational research emerged as a set of practices shaped by the decision at hand, rather than as a single literature-review activity. Stakeholders worked with evidence that was heterogeneous, incompletely reported, unevenly accessible, and sometimes absent from the public record. AI-assisted systems may support retrieval, structuring, and initial comparison, but they cannot resolve all of these challenges and may shift work toward verification and expert interpretation. We therefore argue for interactive systems tailored to the different needs of individual researchers and stakeholder groups, while accounting for the limitations and nuances of the underlying body of evidence.

\subsection{Role- and Practice-Sensitive Tools over a Shared Evidence Layer}

Our findings show that evidence needs vary across stakeholder roles and decision contexts. Basic researchers engaged deeply with the literature to plan experiments, whereas laboratory leaders, license and regulatory reviewers, and animal-welfare stakeholders more often used evidence to justify, assess, or refine proposed work. Evidence-support systems should therefore not assume a single model of evidence work. Instead, they should provide views, levels of detail, and interaction mechanisms suited to the concrete responsibilities and decisions of their intended users.

At the same time, participants within the same stakeholder groups differed in the tools, search strategies, and expertise-driven judgments they applied. Role-sensitive design should therefore not treat professional roles as homogeneous user categories. Systems could allow users to articulate or adjust their search priorities, inclusion criteria, and preferred evidence sources while keeping these choices visible and revisable. This aligns with HCI work showing that stakeholder needs depend not only on formal roles, but also on users' goals, knowledge, tasks, and organizational contexts \cite{sureshBeyondExpertiseRoles2021,hongHumanFactorsModel2020}. Related work on narrative reviews further suggests that interactive systems can help externalize the otherwise implicit strategies that guide evidence discovery and synthesis \cite{fok2025toward}.

Decisions about animal experiments are interdependent across planning, review, and translation. Researchers use prior evidence to select models and procedures and to justify a proposed study. Animal-welfare officers and licensing bodies subsequently assess this justification and may request further evidence, alternative methods, or refinements. In translational research, these choices also affect later assessments of relevance to human disease and clinical development. This is consistent with prior work describing animal-study planning and approval as a collaborative process involving scientific, ethical, and 3R considerations \cite{smithPREPAREGuidelinesPlanning2018}.

Future systems should therefore combine role- and practice-sensitive interfaces with support for coordination across the wider translational process. Clinical decision-support research similarly argues that interconnected decisions can benefit from tools drawing on a consistent evidence source while adapting information to the specific user and decision context \cite{yangHarnessingBiomedicalLiterature2023}. Tailored tools could operate over a partially shared evidence layer, allowing evidence and rationale recorded during study planning to inform later welfare, licensing, and translational assessments.  
Because some regulatory, institutional, and commercial evidence is restricted, such infrastructure may also need to support differentiated access rights.
A recent framework illustrates this direction by integrating animal, in vitro, clinical, pharmacological, and feasibility evidence into living summaries used to support drug prioritization for clinical trials \cite{wong2025systematic}.

\subsection{Supporting Systematic Assessment under Evidence Limitations}

Our findings show that the scale, heterogeneity, and incomplete reporting of animal-research evidence make comprehensive assessment difficult. Participants therefore relied on heuristics such as journal reputation, citation counts, familiar research groups, and prior expertise, despite recognizing their limitations. Similar selective and fragmented citation practices have been observed in broader scholarly work \cite{fangExploringPracticesChallenges2025}. Our findings show how the specific constraints of animal-research evidence can further reinforce reliance on such imperfect judgments.

These limitations strengthen the need for more systematic assessment. Because new animal studies build on an already heterogeneous and sometimes unreliable evidence base, researchers need a clear account of what is known, uncertain, or conflicting before designing further experiments. Structured search, selection, and comparison can support more transparent synthesis across studies by making the evidence and criteria informing expert judgment more explicit~\cite{ioannidisSystematicReviewsBasic2023}. AI-assisted systems could make such assessment more feasible by supporting study identification, structured extraction, and comparison across heterogeneous evidence. They could also help surface where findings converge, conflict, or remain uncertain.

One way to strengthen these capabilities is through semantically aligned evidence representations. AI-assisted systems could help map heterogeneous terminology and extract study characteristics into shared ontologies and standardized relations \cite{soaresMakingScienceComputable2024}. In practice, this could take the form of a knowledge graph connecting studies, models, interventions, and outcomes through shared concepts and relationships, building on broader efforts to structure heterogeneous biomedical knowledge in graph-based representations~\cite{chandak2023building}. In turn, this structured layer could support more consistent comparison and reasoning across studies for both systems and users, while making terminological and methodological differences more visible and facilitating the reuse of curated evidence. However, these connections should not imply that related studies, models, or outcomes are directly comparable, since their translational relevance still requires expert interpretation.  
Initiatives such as the Monarch Initiative illustrate the potential of interoperable semantic resources for this purpose \cite{putman2024monarch}. 

More fundamentally, structuring existing evidence is not enough when relevant findings remain unpublished, inaccessible, or poorly reported.
AI could address part of this problem at earlier stages by checking protocols, preregistrations, and manuscripts for missing methodological and
3R-related information~\cite{wangDevelopmentValidationNatural2020,menkeRigorTransparencyIndex2020,nc3rsDevelopingAITool2023}. Such systems may improve the completeness of future reports, but they cannot recover findings that were never published or make restricted evidence accessible. They must therefore be accompanied by coordinated efforts from researchers, institutions, publishers, and funders to strengthen open science and improve the completeness and transparency of research reporting~\cite{reynoldsReportingTransparencyLaboratory2026,bannach-brownBuildingSynthesisreadyResearch2025,brazilIlluminatingUglySide2024}.

\subsection{Designing for Verifiable, Contextualized, and Collaborative AI Support}

Our findings suggest that participants envisioned AI-assisted evidence tools as first-pass synthesizers rather than replacements for expert judgment. Such systems could reduce the effort required to gather and organize evidence while shifting expert work toward assessing coverage, checking sources, and interpreting findings in context. Participants therefore emphasized the need to inspect original publications, verify extracted information, distinguish reported evidence from generated interpretation, understand uncertainty, and ask follow-up questions.

What counts as sufficient validation, however, depends on the user's role and task. A researcher selecting an animal model may require detailed methodological information, whereas a reviewer may need to trace the evidence supporting a scientific justification or refinement claim. The design and evaluation of evidence tools should therefore reflect users' specific responsibilities, decisions, and validation requirements \cite{liaoConnectingAlgorithmicResearch2022}. This is particularly important given known problems with hallucination, weak source attribution, and the limited performance of LLMs on expert evidence-synthesis tasks \cite{xuNaturalLanguageProcessing2025,solliniHumanResearchersAre2025}.

Participants' positive responses to the dashboard indicate that structured, study-level tables may provide a useful shared starting point. They can offer an overview of the available evidence while preserving access to individual studies before users proceed to task-specific comparison and interpretation. This aligns with evidence-synthesis guidance highlighting the value of tabular displays for communicating detail efficiently, revealing patterns, and supporting traceability \cite{mckenzieSynthesizingPresentingFindings2024}. Semantically organized, interactive evidence maps offer a complementary direction by helping users explore themes and move from higher-level patterns to relevant original sources \cite{mozgaiAcceleratingScopingReviews2024}.

Beyond static evidence displays, our findings also point toward conversational, multi-turn exploration, where users iteratively refine questions and probe why studies differ or how evidence relates to a translational problem. This suggests opportunities for conversational or agent-like evidence assistants that support progressive sensemaking rather than single-query retrieval \cite{skarlinski2024language,gao2025democratizing,fok2025toward}, while preserving source traceability and expert control.

These findings support a human--AI collaboration paradigm in which AI contributes speed and scale in evidence aggregation and organization, while experts retain responsibility for critical assessment and contextual interpretation \cite{thurzoRevisitingRoleReview2025}. Ongoing efforts to develop and validate foundation models for literature mining in clinical research demonstrate the potential of this direction \cite{wangFoundationModelHumanAI2025a}. Our findings extend this direction to preclinical evidence work by identifying the practices, validation needs, and translational challenges that such systems must address. 

This perspective also connects to emerging visions of AI for science, where AI can accelerate literature synthesis, analysis, and scientific discovery while complementing human expertise \cite{asaiSynthesizingScientificLiterature2026,shaoSciSciGPTAdvancingHuman2026}. At the same time, recent work suggests that increased productivity may come with risks such as narrowing the diversity of research topics pursued, underscoring the need for AI systems that augment rather than constrain scientific judgment \cite{haoArtificialIntelligenceTools2026a}. These broader concerns reinforce the importance of examining how such systems shape scientific reasoning, evidence use, and decision-making in real-world research workflows.

Taken together, our findings identify three tensions for future AI-assisted evidence systems: first, they must organize a shared evidence base while accommodating different stakeholder goals and evaluative criteria; second, structure available evidence while making its incompleteness visible; and third, automate parts of evidence work while preserving traceability and opportunities for expert verification. Addressing these tensions could contribute to more efficient and evidence-based drug development while supporting high standards of animal welfare and the 3Rs.

\section{Limitations and Future Work}

\noindent
\textbf{Sample scope and transferability}. We interviewed 13 participants across four stakeholder categories, all based in Switzerland. The sample covered animal research, animal welfare and the 3Rs,
regulatory and ethics, and clinical and drug development perspectives, but did not include practising clinicians or regulators from FDA or EMA contexts. Some participants could fit into more than one stakeholder category, so we grouped them according to their main professional role. 
The findings may be relevant to similar settings. Their transferability to other stakeholder groups or to institutional and regulatory contexts in other countries may be limited. Future studies could therefore include additional professional groups, countries, and institutions.

\noindent
\textbf{Self-reported workflows and first-impression evaluation}. Participants described their literature practices from memory. They may therefore have focused on difficult or memorable experiences and said less about routine activities. Future work could use contextual inquiry or other observational methods to study literature-review practices as they happen. The dashboard was evaluated during one brief think-aloud session. Participants' initial reactions may therefore reflect the limited time available to explore the system. The findings concern perceived usefulness, anticipated use, and information needs rather than sustained adoption, efficiency gains, extraction accuracy, or 3R outcomes. Longer-term studies using real research tasks could examine how the tool fits into evidence-review workflows.

\noindent
\textbf{Scope and future development of the dashboard}. Participants evaluated one version of the dashboard with a specific literature collection, organization, and set of features. Their feedback, therefore, applies mainly to this prototype rather than to AI-assisted evidence tools in general. Future versions should separate individual experiments within papers, include quantitative outcome data and animal-welfare information, support semantic search, link extracted information to the supporting source sentences, and provide clearer information about the scope and update status of the literature collection.

\section{Conclusion}

Evidence work in animal and translational research is not a single
workflow but a set of role- and decision-dependent practices. We
conducted semi-structured interviews with 13 stakeholders to
investigate how evidence is found, assessed, and connected across
animal and translational research, and how participants responded to a
curated evidence dashboard. These practices were constrained by
incomplete reporting, restricted access, unpublished findings, and
difficult cross-study and cross-species comparison. AI-assisted tools
can support retrieval and evidence structuring, but also shift effort
toward verification and expert interpretation.

These findings suggest that future systems should provide role- and practice-sensitive views over shared evidence, expose coverage and uncertainty, preserve source-level traceability, and retain human control over consequential judgments. Rather than replacing expert review, such systems should reduce the effort of gathering and organizing evidence while supporting the verification, contextual interpretation, and judgment that remain with domain experts.


\begin{acks}
We thank all interview participants for sharing their time, experiences, and valuable insights, which formed the foundation of this paper.
\end{acks}

\section*{Ethics and Privacy Statement}

This study was reviewed through our institution's data protection compliance
process and determined not to require submission to the Cantonal Ethics
Committee under the Swiss Human Research Act. Participants provided informed
consent, had the right to withdraw at any time, and permitted audio recording
of the interviews. Direct identifiers were removed, and participants are
reported using pseudonymous identifiers (P01--P13) and broad professional
categories to reduce identification risk. Study data were stored securely on
institutional infrastructure and accessed only by the research team.

\bibliographystyle{ACM-Reference-Format}
\bibliography{sample-base}

@String{Computing = "Computing" }

@String{Computer = "{IEEE} Computer" }

@String{Academic = "Academic Press" }

@String{Springer = "Springer-Verlag" }

@article{ineichen2023data,
  title     = {From data deluge to publomics: How AI can transform animal research},
  author    = {Ineichen, Benjamin V. and Rosso, Marianna and Macleod, Malcolm R.},
  journal   = {Lab Animal},
  volume    = {52},
  number    = {10},
  pages     = {213--214},
  year      = {2023},
  publisher = {Nature Publishing Group},
  doi       = {10.1038/s41684-023-01256-4}
}

@article{ineichenAnalysisAnimaltohumanTranslation2024,
  title = {Analysis of Animal-to-Human Translation Shows That Only 5\% of Animal-Tested Therapeutic Interventions Obtain Regulatory Approval for Human Applications},
  author = {Ineichen, Benjamin V. and Furrer, Eva and Gr{\"u}ninger, Servan L. and Z{\"u}rrer, Wolfgang E. and Macleod, Malcolm R.},
  year = {2024},
  journal = {PLOS Biology},
  volume = {22},
  number = {6},
  pages = {e3002667},
  publisher = {Public Library of Science},
  issn = {1545-7885},
  doi = {10.1371/journal.pbio.3002667},
  urldate = {2024-06-27},
  langid = {english},

}

@article{comanici2025gemini,
  title={Gemini 2.5: Pushing the frontier with advanced reasoning, multimodality, long context, and next generation agentic capabilities},
  author={Comanici, Gheorghe and Bieber, Eric and Schaekermann, Mike and Pasupat, Ice and Sachdeva, Noveen and Dhillon, Inderjit and Blistein, Marcel and Ram, Ori and Zhang, Dan and Rosen, Evan and others},
  journal={arXiv preprint arXiv:2507.06261},
  year={2025}
}

@misc{benchsci2026emet,
  author       = {{BenchSci}},
  title        = {EMET Documentation},
  year         = {2026},
  howpublished = {\url{https://www.benchsci.com/help}},
  note         = {Accessed: 2026-09-07}
}

@article{kimFactorsAffectingSuccess2023a,
  title = {Factors {{Affecting Success}} of {{New Drug Clinical Trials}}},
  author = {Kim, Eungdo and Yang, Jaehoon and Park, Sungjin and Shin, Kwangsoo},
  year = 2023,
  month = jul,
  journal = {Therapeutic Innovation \& Regulatory Science},
  volume = {57},
  number = {4},
  pages = {737--750},
  issn = {2168-4804},
  doi = {10.1007/s43441-023-00509-1},
  urldate = {2024-11-27},
  langid = {english},
}

@article{bespalovFailedTrialsCentral2016a,
  title = {Failed Trials for Central Nervous System Disorders Do Not Necessarily Invalidate Preclinical Models and Drug Targets},
  author = {Bespalov, Anton and Steckler, Thomas and Altevogt, Bruce and Koustova, Elena and Skolnick, Phil and Deaver, Daniel and Millan, Mark J. and Bastlund, Jesper F. and Doller, Dario and Witkin, Jeffrey and Moser, Paul and O'Donnell, Patricio and Ebert, Ulrich and Geyer, Mark A. and Prinssen, Eric and Ballard, Theresa and Macleod, Malcolm},
  year = 2016,
  month = jul,
  journal = {Nature Reviews Drug Discovery},
  volume = {15},
  number = {7},
  pages = {516--516},
  issn = {1474-1776, 1474-1784},
  doi = {10.1038/nrd.2016.88},
  urldate = {2025-06-11},
  langid = {english},

}

@article{ferreiraLevellingTranslationalGap2020,
  title = {Levelling the {{Translational Gap}} for {{Animal}} to {{Human Efficacy Data}}},
  author = {Ferreira, Guilherme S. and {Veening-Griffioen}, D{\'e}sir{\'e}e H. and Boon, Wouter P. C. and Moors, Ellen H. M. and {van Meer}, Peter J. K.},
  year = 2020,
  month = jul,
  journal = {Animals},
  volume = {10},
  number = {7},
  pages = {1199},
  publisher = {Multidisciplinary Digital Publishing Institute},
  issn = {2076-2615},
  doi = {10.3390/ani10071199},
  urldate = {2025-01-16},
  copyright = {http://creativecommons.org/licenses/by/3.0/},
  langid = {english},
}

@article{borahAnalysisTimeWorkers2017a,
  title = {Analysis of the Time and Workers Needed to Conduct Systematic Reviews of Medical Interventions Using Data from the {{PROSPERO}} Registry},
  author = {Borah, Rohit and Brown, Andrew W and Capers, Patrice L and Kaiser, Kathryn A},
  year = 2017,
  month = feb,
  journal = {BMJ Open},
  volume = {7},
  number = {2},
  pages = {e012545},
  issn = {2044-6055, 2044-6055},
  doi = {10.1136/bmjopen-2016-012545},
  urldate = {2026-09-08},
  langid = {english},
}

@article{bugajskaHowLongDoes2025,
  title = {How Long Does It Take to Complete and Publish a Systematic Review of Animal Studies?},
  author = {Bugajska, Julia Victoria and Hild, Bernard Friedrich and Br{\"u}schweiler, David and Meier, Enrico Daniele and {Bannach-Brown}, Alexandra and Wever, Kimberley Elaine and Ineichen, Benjamin Victor},
  year = 2025,
  month = oct,
  journal = {BMC Medical Research Methodology},
  volume = {25},
  number = {1},
  pages = {226},
  issn = {1471-2288},
  doi = {10.1186/s12874-025-02672-5},
  urldate = {2026-02-25},
  langid = {english},
}

@article{cleoUsabilityAcceptabilityFour2019,
  title = {Usability and Acceptability of Four Systematic Review Automation Software Packages: A Mixed Method Design},
  shorttitle = {Usability and Acceptability of Four Systematic Review Automation Software Packages},
  author = {Cleo, Gina and Scott, Anna Mae and Islam, Farhana and Julien, Blair and Beller, Elaine},
  year = 2019,
  month = jun,
  journal = {Systematic Reviews},
  volume = {8},
  number = {1},
  pages = {145},
  issn = {2046-4053},
  doi = {10.1186/s13643-019-1069-6},
  urldate = {2026-09-08},
  langid = {english},
}

@article{SoftwareToolsSystematic2024,
  title = {Software Tools for Systematic Review Literature Screening and Data Extraction: {{Qualitative}} User Experiences from Succinct Formal Tests},
  shorttitle = {Software Tools for Systematic Review Literature Screening and Data Extraction},
  author = {Leenaars, Cathalijn HC and Stafleu, Frans and Andre, Bleich},
  year = 2024,
  journal = {ALTEX},
  issn = {1868596X},
  doi = {10.14573/altex.2409251},
  urldate = {2026-09-08},
  langid = {english}
}

@article{chandak2023building,
  title={Building a knowledge graph to enable precision medicine},
  author={Chandak, Payal and Huang, Kexin and Zitnik, Marinka},
  journal={Scientific Data},
  volume={10},
  number={1},
  pages={67},
  year={2023},
  publisher={Nature Publishing Group UK London},
  doi={10.1038/s41597-023-01960-3}
}

@article{minozzi2020revised,
  title={The revised Cochrane risk of bias tool for randomized trials (RoB 2) showed low interrater reliability and challenges in its application},
  author={Minozzi, Silvia and Cinquini, Michela and Gianola, Silvia and Gonzalez-Lorenzo, Marien and Banzi, Rita},
  journal={Journal of clinical epidemiology},
  volume={126},
  pages={37--44},
  year={2020},
  publisher={Elsevier}
}

@article{wurbelMore3RsImportance2017,
  title = {{{More}} than {{3Rs}}: The Importance of Scientific Validity for Harm-Benefit Analysis of Animal Research},
  shorttitle = {More than {{3Rs}}},
  author = {W{\"u}rbel, Hanno},
  year = 2017,
  month = apr,
  journal = {Lab Animal},
  volume = {46},
  number = {4},
  pages = {164--166},
  issn = {0093-7355, 1548-4475},
  doi = {10.1038/laban.1220},
  urldate = {2025-06-11},
  langid = {english},
}

@book{russell1959principles,
  author    = {Russell, William Moy Stratton and Burch, Rex Leonard},
  title     = {The Principles of Humane Experimental Technique},
  year      = {1959},
  publisher = {Methuen},
  address   = {London}
}

@article{sousaLandscapeArtificialIntelligence2026,
  title = {{{The}} Landscape of Artificial Intelligence Tools and Platforms for Evidence Synthesis: A Scoping Review},
  shorttitle = {The Landscape of Artificial Intelligence Tools and Platforms for Evidence Synthesis},
  author = {Sousa, M. Sharmila A. and Peiris, Sasha and Figueir{\'o}, Mabel F. and Haby, Michelle M. and Baraldi, Ana Cyntia and Reveiz, Ludovic and Souza, Jo{\~a}o Paulo},
  year = 2026,
  month = feb,
  journal = {Systematic Reviews},
  volume = {15},
  number = {1},
  pages = {82},
  issn = {2046-4053},
  doi = {10.1186/s13643-025-02842-y},
  urldate = {2026-03-19},
  langid = {english},

}

@article{braunReflectingReflexiveThematic2019,
  title = {Reflecting on Reflexive Thematic Analysis},
  author = {Braun, Virginia and Clarke, Victoria},
  year = 2019,
  month = aug,
  journal = {Qualitative Research in Sport, Exercise and Health},
  volume = {11},
  number = {4},
  pages = {589--597},
  issn = {2159-676X, 2159-6778},
  doi = {10.1080/2159676X.2019.1628806},
  urldate = {2026-06-08},
  langid = {english},
}

@article{braunUsingThematicAnalysis2006,
  title = {Using Thematic Analysis in Psychology},
  author = {Braun, Virginia and Clarke, Victoria},
  year = 2006,
  month = jan,
  journal = {Qualitative Research in Psychology},
  volume = {3},
  number = {2},
  pages = {77--101},
  issn = {1478-0887, 1478-0895},
  doi = {10.1191/1478088706qp063oa},
  urldate = {2026-06-08},
  langid = {english},
}

@software{curtain2025qualcoder,
  author  = {Curtain, Colin and Dröge, Kai and Missaghieh--Poncet, Katharina and Salomón, Luis},
  title   = {QualCoder},
  version = {3.8.2},
  year    = {2025},
  url     = {https://github.com/ccbogel/QualCoder/releases/tag/3.8.2},
  note    = {Computer software}
}

@article{eger2025transforming,
  title={Transforming science with large language models: A survey on ai-assisted scientific discovery, experimentation, content generation, and evaluation},
  author={Eger, Steffen and Cao, Yong and D'Souza, Jennifer and Geiger, Andreas and Greisinger, Christian and Gross, Stephanie and Hou, Yufang and Krenn, Brigitte and Lauscher, Anne and Li, Yizhi and others},
  journal={arXiv preprint arXiv:2502.05151},
  year={2025}
}

@inproceedings{mikriukov2025ai,
  title={AI tools for automating systematic literature reviews},
  author={Mikriukov, Andrei and Senokosov, Artsiom and Succi, Giancarlo and Tormasov, Alexander and Plaksin, Yaroslav and Trofimova, Ekaterina and Sitnikov, Vladimir},
  booktitle={Proceedings of the 2025 International Conference on Software Engineering and Computer Applications},
  pages={25--30},
  year={2025}
}

@article{van2021open,
  title={An open source machine learning framework for efficient and transparent systematic reviews},
  author={Van De Schoot, Rens and De Bruin, Jonathan and Schram, Raoul and Zahedi, Parisa and De Boer, Jan and Weijdema, Felix and Kramer, Bianca and Huijts, Martijn and Hoogerwerf, Maarten and Ferdinands, Gerbrich and others},
  journal={Nature machine intelligence},
  volume={3},
  number={2},
  pages={125--133},
  year={2021},
  publisher={Nature Publishing Group UK London}
}

@article{skarlinski2024language,
  title={Language agents achieve superhuman synthesis of scientific knowledge},
  author={Skarlinski, Michael D and Cox, Sam and Laurent, Jon M and Braza, James D and Hinks, Michaela and Hammerling, Michael J and Ponnapati, Manvitha and Rodriques, Samuel G and White, Andrew D},
  journal={arXiv preprint arXiv:2409.13740},
  year={2024}
}

@inproceedings{doneva2025preclinie,
  title={PreClinIE: An Annotated Corpus for Information Extraction in Preclinical Studies},
  author={Doneva, Simona and Hubarava, Hanna and Haervelid, Pia and Z{\"u}rrer, Wolfgang and Bugajska, Julia and Hild, Bernard and Br{\"u}schweiler, David and Schneider, Gerold and Ellendorff, Tilia and Ineichen, Benjamin},
  booktitle={Proceedings of the 24th Workshop on Biomedical Language Processing},
  pages={74--87},
  year={2025}
}

@article{asai2026synthesizing,
  title={Synthesizing scientific literature with retrieval-augmented language models},
  author={Asai, Akari and He, Jacqueline and Shao, Rulin and Shi, Weijia and Singh, Amanpreet and Chang, Joseph Chee and Lo, Kyle and Soldaini, Luca and Feldman, Sergey and D’Arcy, Mike and others},
  journal={Nature},
  pages={1--7},
  year={2026},
  publisher={Nature Publishing Group UK London}
}

@article{peeters2025evaluation,
  title={Evaluation of SURUS: a named entity recognition NLP system to extract knowledge from interventional study records},
  author={Peeters, Casper and Vijverberg, Koen and Pouwer, Marianne and Westerman, Bart and Boot, Maikel and Verberne, Suzan},
  journal={BMC Medical Research Methodology},
  volume={25},
  number={1},
  pages={184},
  year={2025},
  publisher={Springer}
}

@article{zurrer2024steed,
  title={STEED: A data mining tool for automated extraction of experimental parameters and risk of bias items from in vivo publications},
  author={Zurrer, Wolfgang Emanuel and Cannon, Amelia Elaine and Ewing, Ewoud and Br{\"u}schweiler, David and Bugajska, Julia and Hild, Bernard Friedrich and Rosso, Marianna and Reich, Daniel Salo and Ineichen, Benjamin Victor},
  journal={PloS one},
  volume={19},
  number={11},
  pages={e0311358},
  year={2024},
  publisher={Public Library of Science San Francisco, CA USA}
}

@article{gao2025democratizing,
  title={Democratizing AI scientists using ToolUniverse},
  author={Gao, Shanghua and Zhu, Richard and Sui, Pengwei and Kong, Zhenglun and Aldogom, Sufian and Huang, Yepeng and Noori, Ayush and Shamji, Reza and Parvataneni, Krishna and Tsiligkaridis, Theodoros and others},
  journal={arXiv preprint arXiv:2509.23426},
  year={2025}
}

@inproceedings{kambhamettu2025traceable,
  title={Traceable Texts and Their Effects: A Study of Summary-Source Links in AI-Generated Summaries},
  author={Kambhamettu, Hita and Flores, Jamie and Head, Andrew},
  booktitle={Proceedings of the Extended Abstracts of the CHI Conference on Human Factors in Computing Systems},
  pages={1--7},
  year={2025}
}

@inproceedings{schemmer2023appropriate,
  title={Appropriate reliance on AI advice: Conceptualization and the effect of explanations},
  author={Schemmer, Max and Kuehl, Niklas and Benz, Carina and Bartos, Andrea and Satzger, Gerhard},
  booktitle={Proceedings of the 28th International Conference on Intelligent User Interfaces},
  pages={410--422},
  year={2023}
}

@inproceedings{nushi2019guidelines,
  title={Guidelines for human-AI interaction},
  author={Nushi, Besmira and Iqbal, Shamsi T and Bennett, Paul and Teevan, Jaime and Horvitz, Eric and Weld, Dan and Kikin-Gil, Ruth and Amershi, Saleema and Collisson, Penny and Vorvoreanu, Mihaela and others},
  booktitle={Proceedings of the 2019 chi conference on human factors in computing systems},
  year={2019}
}

@inproceedings{liao2020questioning,
  title={Questioning the AI: informing design practices for explainable AI user experiences},
  author={Liao, Q Vera and Gruen, Daniel and Miller, Sarah},
  booktitle={Proceedings of the 2020 CHI conference on human factors in computing systems},
  pages={1--15},
  year={2020}
}

@inproceedings{jacobs2021designing,
  title={Designing AI for trust and collaboration in time-constrained medical decisions: a sociotechnical lens},
  author={Jacobs, Maia and He, Jeffrey and F. Pradier, Melanie and Lam, Barbara and Ahn, Andrew C and McCoy, Thomas H and Perlis, Roy H and Doshi-Velez, Finale and Gajos, Krzysztof Z},
  booktitle={Proceedings of the 2021 chi conference on human factors in computing systems},
  pages={1--14},
  year={2021}
}

@article{zhang2024leveraging,
  title={Leveraging generative AI for clinical evidence synthesis needs to ensure trustworthiness},
  author={Zhang, Gongbo and Jin, Qiao and McInerney, Denis Jered and Chen, Yong and Wang, Fei and Cole, Curtis L and Yang, Qian and Wang, Yanshan and Malin, Bradley A and Peleg, Mor and others},
  journal={Journal of biomedical informatics},
  volume={153},
  pages={104640},
  year={2024},
  publisher={Elsevier}
}

@article{bach2024systematic,
  title={A systematic literature review of user trust in AI-enabled systems: An HCI perspective},
  author={Bach, Tita Alissa and Khan, Amna and Hallock, Harry and Beltr{\~a}o, Gabriela and Sousa, Sonia},
  journal={International Journal of Human--Computer Interaction},
  volume={40},
  number={5},
  pages={1251--1266},
  year={2024},
  publisher={Taylor \& Francis}
}

@article{devriesUsefulnessSystematicReviews2014b,
  title = {{{The Usefulness}} of {{Systematic Reviews}} of {{Animal Experiments}} for the {{Design}} of {{Preclinical}} and {{Clinical Studies}}},
  author = {{de Vries}, Rob B. M. and Wever, Kimberley E. and Avey, Marc T. and Stephens, Martin L. and Sena, Emily S. and Leenaars, Marlies},
  year = 2014,
  month = dec,
  journal = {ILAR Journal},
  volume = {55},
  number = {3},
  pages = {427--437},
  issn = {1930-6180},
  doi = {10.1093/ilar/ilu043},
  urldate = {2026-03-30},
}

@article{ineichenSystematicReviewMetaanalysis2024a,
  title = {Systematic Review and Meta-Analysis of Preclinical Studies},
  author = {Ineichen, Benjamin Victor and Held, Ulrike and Salanti, Georgia and Macleod, Malcolm Robert and Wever, Kimberley Elaine},
  year = 2024,
  month = oct,
  journal = {Nature Reviews Methods Primers},
  volume = {4},
  number = {1},
  pages = {72},
  publisher = {Nature Publishing Group},
  issn = {2662-8449},
  doi = {10.1038/s43586-024-00347-x},
  urldate = {2026-03-13},
  copyright = {2024 Springer Nature Limited},
  langid = {english},
}

@article{marshallTowardSystematicReviewAutomation2019,
  title={Toward systematic review automation: a practical guide to using machine learning tools in research synthesis},
  author={Marshall, Christopher and Wallace, Byron C},
  journal={Systematic Reviews},
  year={2019},
  volume={8},
  pages={163},
  doi={10.1186/s13643-019-1074-9}
}

@article{pistollatoEvaluatingTranslationalValue2026,
  title = {Evaluating the Translational Value of Preclinical Models: {{Available}} Tools and Frameworks, Challenges and Strategies},
  shorttitle = {Evaluating the Translational Value of Preclinical Models},
  author = {Pistollato, Francesca and Furtmann, Fabia and Straccia, Marco and Avey, Marc and Azilagbetor, David Mawufemor and Camp, Celean and Delaney, Conor and Ferreira, Guilherme S. and {Garcia-Bermejo}, Maria Laura and Gastaldello, Annalisa and Gurusamy, Kurinchi and Holden, Laura and Kimmelman, Jonathan and Lohse, Simon and Marigliani, Bianca and Menon, Julia M.L. and {Ritskes-Hoitinga}, Merel and Sarasija, Shaarika and Tagle, Danilo and Tripodi, Ignacio J. and Turner, Jan and Wehling, Martin and Costantino, Helder},
  year = 2026,
  month = jan,
  journal = {Alternatives to Laboratory Animals},
  volume = {54},
  number = {1},
  pages = {10--27},
  publisher = {SAGE Publications Ltd STM},
  issn = {0261-1929},
  doi = {10.1177/02611929251398821},
  urldate = {2026-01-22},
  langid = {english},
}

@article{bannach-brownBuildingSynthesisreadyResearch2025,
  title = {{{Building}} a Synthesis-Ready Research Ecosystem: Fostering Collaboration and Open Science to Accelerate Biomedical Translation},
  shorttitle = {Building a Synthesis-Ready Research Ecosystem},
  author = {{Bannach-Brown}, Alexandra and Rackoll, Torsten and Macleod, Malcolm R. and McCann, Sarah K.},
  year = 2025,
  month = mar,
  journal = {BMC Medical Research Methodology},
  volume = {25},
  number = {1},
  pages = {66},
  issn = {1471-2288},
  doi = {10.1186/s12874-025-02524-2},
  urldate = {2026-02-26},
  langid = {english},
}

@article{hairConnectingDotsNeuroscience2025a,
  title = {{{Connecting}} the Dots in Neuroscience Research: {{The}} Future of Evidence Synthesis},
  shorttitle = {Connecting the Dots in Neuroscience Research},
  author = {Hair, Kaitlyn and {Arroyo-Araujo}, Mar{\'i}a and Vojvodic, Sofija and Economou, Maria and Wong, Charis and Tinsdeall, Francesca and Smith, Sean and Rackoll, Torsten and Sena, Emily S. and McCann, Sarah K.},
  year = 2025,
  month = feb,
  journal = {Experimental Neurology},
  volume = {384},
  pages = {115047},
  issn = {0014-4886},
  doi = {10.1016/j.expneurol.2024.115047},
  urldate = {2026-02-08},
}

@article{riazFutureEvidenceSynthesis2024a,
  title = {{{Future}} of {{Evidence Synthesis}}: {{Automated}}, {{Living}}, and {{Interactive Systematic Reviews}} and {{Meta-analyses}}},
  shorttitle = {Future of {{Evidence Synthesis}}},
  author = {Riaz, Irbaz Bin and Naqvi, Syed Arsalan Ahmed and Hasan, Bashar and Murad, Mohammad Hassan},
  year = 2024,
  month = jun,
  journal = {Mayo Clinic Proceedings: Digital Health},
  volume = {2},
  number = {3},
  pages = {361--365},
  issn = {2949-7612},
  doi = {10.1016/j.mcpdig.2024.05.023},
  urldate = {2026-02-04},
  pmcid = {PMC11975841},
  pmid = {40206128},
}

@inproceedings{duckFindingNeedlesDocument2025b,
  title = {Finding {{Needles}} in {{Document Haystacks}}: {{Augmenting Serendipitous Claim Retrieval Workflows}}},
  shorttitle = {Finding {{Needles}} in {{Document Haystacks}}},
  booktitle = {Proceedings of the 2025 {{CHI Conference}} on {{Human Factors}} in {{Computing Systems}}},
  author = {D{\"u}ck, Moritz and Holter, Steffen and Chan, Robin Shing Moon and Sevastjanova, Rita and {El-Assady}, Mennatallah},
  year = 2025,
  month = apr,
  series = {{{CHI}} '25},
  pages = {1--17},
  publisher = {Association for Computing Machinery},
  address = {New York, NY, USA},
  doi = {10.1145/3706598.3713715},
  urldate = {2026-09-08},
  isbn = {979-8-4007-1394-1}
}

@article{Doneva2026,
  author  = {Doneva, Simona Emilova and Sick, Beate and Ellendorff, Tilia R. and Goldman, Jean-Philippe and Reich, Daniel S. and Hemkens, Lars G. and Schneider, Gerold and Ineichen, Benjamin Victor},
  title   = {Design, reporting, and research focus of neurology and psychiatry drug trials (2020--2023): a registry-based meta-epidemiological study using {ClinicalTrials.gov} and {NLP}},
  journal = {Trials},
  year    = {2026},
  doi     = {10.1186/s13063-026-09909-8},
  url     = {https://doi.org/10.1186/s13063-026-09909-8},
  issn    = {1745-6215}
}

@misc{miro2026,
  author       = {{Miro}},
  title        = {Miro: Visual Workspace for Innovation},
  year         = {2026},
  howpublished = {\url{https://miro.com/}},
  note         = {Online collaborative whiteboard}
}

@article{asaiSynthesizingScientificLiterature2026,
  title = {Synthesizing Scientific Literature with Retrieval-Augmented Language Models},
  author = {Asai, Akari and He, Jacqueline and Shao, Rulin and Shi, Weijia and Singh, Amanpreet and Chang, Joseph Chee and Lo, Kyle and Soldaini, Luca and Feldman, Sergey and D'Arcy, Mike and Wadden, David and Latzke, Matt and Sparks, Jenna and Hwang, Jena D. and Kishore, Varsha and Tian, Minyang and Ji, Pan and Liu, Shengyan and Tong, Hao and Wu, Bohao and Xiong, Yanyu and Zettlemoyer, Luke and Neubig, Graham and Weld, Daniel S. and Downey, Doug and Yih, Wen-tau and Koh, Pang Wei and Hajishirzi, Hannaneh},
  year = 2026,
  month = feb,
  journal = {Nature},
  volume = {650},
  number = {8103},
  pages = {857--863},
  publisher = {Nature Publishing Group},
  issn = {1476-4687},
  doi = {10.1038/s41586-025-10072-4},
  urldate = {2026-09-09},
  copyright = {2026 The Author(s)},
  langid = {english}
}

@article{haoArtificialIntelligenceTools2026a,
  title = {Artificial Intelligence Tools Expand Scientists' Impact but Contract Science's Focus},
  author = {Hao, Qianyue and Xu, Fengli and Li, Yong and Evans, James},
  year = 2026,
  month = jan,
  journal = {Nature},
  volume = {649},
  number = {8099},
  pages = {1237--1243},
  publisher = {Nature Publishing Group},
  issn = {1476-4687},
  doi = {10.1038/s41586-025-09922-y},
  urldate = {2026-09-09},
  copyright = {2026 The Author(s), under exclusive licence to Springer Nature Limited},
  langid = {english}
}

@article{shaoSciSciGPTAdvancingHuman2026,
  title = {{{SciSciGPT}}: Advancing Human--{{AI}} Collaboration in the Science of Science},
  shorttitle = {{{SciSciGPT}}},
  author = {Shao, Erzhuo and Wang, Yifang and Qian, Yifan and Pan, Zhenyu and Liu, Han and Wang, Dashun},
  year = 2026,
  month = mar,
  journal = {Nature Computational Science},
  volume = {6},
  number = {3},
  pages = {301--315},
  publisher = {Nature Publishing Group},
  issn = {2662-8457},
  doi = {10.1038/s43588-025-00906-6},
  urldate = {2026-09-09},
  copyright = {2025 The Author(s)},
  langid = {english}
}

@inproceedings{wangEvaluatingLargeLanguage2024,
  title = {Evaluating {{Large Language Models}} on {{Academic Literature Understanding}} and {{Review}}: {{An Empirical Study}} among {{Early-stage Scholars}}},
  shorttitle = {Evaluating {{Large Language Models}} on {{Academic Literature Understanding}} and {{Review}}},
  booktitle = {Proceedings of the 2024 {{CHI Conference}} on {{Human Factors}} in {{Computing Systems}}},
  author = {Wang, Jiyao and Hu, Haolong and Wang, Zuyuan and Yan, Song and Sheng, Youyu and He, Dengbo},
  year = 2024,
  month = may,
  series = {{{CHI}} '24},
  pages = {1--18},
  publisher = {Association for Computing Machinery},
  address = {New York, NY, USA},
  doi = {10.1145/3613904.3641917},
  urldate = {2026-09-08},
  isbn = {979-8-4007-0330-0}
}

@article{ioannidisSystematicReviewsBasic2023,
  title = {Systematic Reviews for Basic Scientists: A Different Beast},
  shorttitle = {Systematic Reviews for Basic Scientists},
  author = {Ioannidis, John P. A.},
  year = 2023,
  month = jan,
  journal = {Physiological Reviews},
  volume = {103},
  number = {1},
  pages = {1--5},
  publisher = {American Physiological Society},
  issn = {0031-9333},
  doi = {10.1152/physrev.00028.2022},
  urldate = {2026-03-13},
}

@article{cookSystematicReviewsSynthesis1997,
  title = {Systematic {{Reviews}}: {{Synthesis}} of {{Best Evidence}} for {{Clinical Decisions}}},
  shorttitle = {Systematic {{Reviews}}},
  author = {Cook, Deborah J. and Mulrow, Cynthia D. and Haynes, R. Brian},
  year = 1997,
  month = mar,
  journal = {Annals of Internal Medicine},
  volume = {126},
  number = {5},
  pages = {376--380},
  publisher = {American College of Physicians},
  issn = {0003-4819},
  doi = {10.7326/0003-4819-126-5-199703010-00006},
  urldate = {2026-07-28},
}

@article{smithPREPAREGuidelinesPlanning2018,
  author  = {Smith, Adrian J. and Clutton, R. Eddie and Lilley, Elliot and Hansen, Kristine E. A. and Brattelid, Trond},
  title   = {{PREPARE}: Guidelines for Planning Animal Research and Testing},
  journal = {Laboratory Animals},
  year    = {2018},
  volume  = {52},
  number  = {2},
  pages   = {135--141},
  doi     = {10.1177/0023677217724823}
}

@inproceedings{yangHarnessingBiomedicalLiterature2023,
  title = {Harnessing {{Biomedical Literature}} to {{Calibrate Clinicians}}' {{Trust}} in {{AI Decision Support Systems}}},
  booktitle = {Proceedings of the 2023 {{CHI Conference}} on {{Human Factors}} in {{Computing Systems}}},
  author = {Yang, Qian and Hao, Yuexing and Quan, Kexin and Yang, Stephen and Zhao, Yiran and Kuleshov, Volodymyr and Wang, Fei},
  year = 2023,
  month = apr,
  series = {{{CHI}} '23},
  pages = {1--14},
  publisher = {Association for Computing Machinery},
  address = {New York, NY, USA},
  doi = {10.1145/3544548.3581393},
  urldate = {2026-07-29},
  isbn = {978-1-4503-9421-5},
}

@inproceedings{liaoConnectingAlgorithmicResearch2022,
  author    = {Liao, Q. Vera and Zhang, Yunfeng and Luss, Ronny and
               Doshi-Velez, Finale and Dhurandhar, Amit},
  title     = {Connecting Algorithmic Research and Usage Contexts:
               A Perspective of Contextualized Evaluation for Explainable AI},
  booktitle = {Proceedings of the AAAI Conference on Human Computation and Crowdsourcing},
  year      = {2022},
  volume    = {10},
  number    = {1},
  pages     = {147--159},
  doi       = {10.1609/hcomp.v10i1.21995}
}

@inproceedings{fok2025toward,
author = {Fok, Raymond and Siu, Alexa and Weld, Daniel S.},
title = {Toward Living Narrative Reviews: An Empirical Study of the Processes and Challenges in Updating Survey Articles in Computing Research},
year = {2025},
isbn = {9798400713941},
publisher = {Association for Computing Machinery},
address = {New York, NY, USA},
url = {https://doi.org/10.1145/3706598.3714047},
doi = {10.1145/3706598.3714047},
booktitle = {Proceedings of the 2025 CHI Conference on Human Factors in Computing Systems},
articleno = {1075},
numpages = {10},
location = {
},
series = {CHI '25}
}

@article{menkeRigorTransparencyIndex2020,
  author  = {Menke, Joe and Roelandse, Martijn and Ozyurt, Burak and
             Martone, Maryann and Bandrowski, Anita},
  title   = {The Rigor and Transparency Index Quality Metric for Assessing
             Biological and Medical Science Methods},
  journal = {iScience},
  year    = {2020},
  volume  = {23},
  number  = {11},
  pages   = {101698},
  doi     = {10.1016/j.isci.2020.101698}
}

@article{wangDevelopmentValidationNatural2020,
  title = {Development and Validation of a Natural Language Processing Tool to Generate the {{CONSORT}} Reporting Checklist for Randomized Clinical Trials},
  author = {Wang, Fan and Schilsky, Richard L. and Page, David and Califf, Robert M. and Cheung, Kei and Wang, Xiaofei and Pang, Herbert},
  year = 2020,
  month = oct,
  journal = {JAMA Network Open},
  volume = {3},
  number = {10},
  pages = {e2014661-e2014661},
  issn = {2574-3805},
  doi = {10.1001/jamanetworkopen.2020.14661}
}

@article{putman2024monarch,
  title={The Monarch Initiative in 2024: an analytic platform integrating phenotypes, genes and diseases across species},
  author={Putman, Tim E and Schaper, Kevin and Matentzoglu, Nicolas and Rubinetti, Vincent P and Alquaddoomi, Faisal S and Cox, Corey and Caufield, J Harry and Elsarboukh, Glass and Gehrke, Sarah and Hegde, Harshad and others},
  journal={Nucleic acids research},
  volume={52},
  number={D1},
  pages={D938--D949},
  year={2024},
  publisher={Oxford University Press},
  doi = {https://doi.org/10.1093/nar/gkad1082}
}

@misc{nc3rsDevelopingAITool2023,
  author       = {{National Centre for the Replacement, Refinement and Reduction of Animals in Research}},
  title        = {Developing an {AI} Tool to Check {ARRIVE} Compliance},
  year         = {2023},
  month        = may,
  day          = {11},
  howpublished = {\url{https://arriveguidelines.org/news/arrive-compliance-checker}},
  note         = {Accessed 30 July 2026}
}

@article{reynoldsReportingTransparencyLaboratory2026,
  title = {Reporting and Transparency in Laboratory Animal Science: {{Past}} Achievements, Present Concerns and Future Directions},
  shorttitle = {Reporting and Transparency in Laboratory Animal Science},
  author = {Reynolds, Penny S. and Pearl, Esther J.},
  year = 2026,
  month = jul,
  journal = {Laboratory Animals},
  pages = {00236772261462343},
  publisher = {SAGE Publications},
  issn = {0023-6772},
  doi = {10.1177/00236772261462343},
  urldate = {2026-07-30},
  langid = {english},

}

@inproceedings{sureshBeyondExpertiseRoles2021,
  author    = {Suresh, Harini and Gomez, Steven R. and Nam, Kevin K. and
               Satyanarayan, Arvind},
  title     = {Beyond Expertise and Roles: A Framework to Characterize the
               Stakeholders of Interpretable Machine Learning and Their Needs},
  booktitle = {Proceedings of the 2021 CHI Conference on Human Factors in
               Computing Systems},
  year      = {2021},
  publisher = {Association for Computing Machinery},
  address   = {New York, NY, USA},
  articleno = {45},
  numpages  = {16},
  doi       = {10.1145/3411764.3445088}
}

@article{hongHumanFactorsModel2020,
  author    = {Hong, Sungsoo Ray and Hullman, Jessica and Bertini, Enrico},
  title     = {Human Factors in Model Interpretability: Industry Practices,
               Challenges, and Needs},
  journal   = {Proceedings of the ACM on Human-Computer Interaction},
  year      = {2020},
  volume    = {4},
  number    = {CSCW1},
  articleno = {68},
  numpages  = {26},
  doi       = {10.1145/3392878}
}

@article{wong2025systematic,
  author  = {Wong, Charis and Cardinali, Alessandra and Liao, Jing and
             Selvaraj, Bhuvaneish T. and Baxter, Paul and Carter, Roderick N.
             and Longden, James and Graham, Rebecca E. and Dakin, Rachel S.
             and Pal, Suvankar and others},
  title   = {Systematic Living Evidence for Clinical Trials ({SyLECT}):
             A Data-Driven Framework for Drug Selection in Clinical Trials
             in Motor Neuron Disease},
  journal = {medRxiv},
  year    = {2025},
  doi     = {10.1101/2025.03.09.25323612},
  note    = {Preprint}
}

@article{brazilIlluminatingUglySide2024,
  title = {Illuminating `the Ugly Side of Science': Fresh Incentives for Reporting Negative Results},
  shorttitle = {Illuminating `the Ugly Side of Science'},
  author = {Brazil, Rachel},
  year = 2024,
  month = may,
  journal = {Nature},
  publisher = {Nature Publishing Group},
  doi = {10.1038/d41586-024-01389-7},
  urldate = {2026-07-30},
  copyright = {2024 Springer Nature Limited},
  langid = {english},
}

@inproceedings{fangExploringPracticesChallenges2025,
  title = {Exploring {{Practices}}, {{Challenges}}, and {{Design Implications}} for {{Citation Foraging}}, {{Management}}, and {{Synthesis}}},
  booktitle = {Proceedings of the {{Extended Abstracts}} of the {{CHI Conference}} on {{Human Factors}} in {{Computing Systems}}},
  author = {Fang, Xinrui and Xu, Anran and Malacria, Sylvain and Yatani, Koji},
  year = 2025,
  month = apr,
  series = {{{CHI EA}} '25},
  pages = {1--8},
  publisher = {Association for Computing Machinery},
  address = {New York, NY, USA},
  doi = {10.1145/3706599.3719883},
  urldate = {2026-07-15},
  isbn = {979-8-4007-1395-8},
}

@inproceedings{xuNaturalLanguageProcessing2025,
  title = {Natural {{Language Processing}} in {{Support}} of {{Evidence-based Medicine}}: {{A Scoping Review}}},
  shorttitle = {Natural {{Language Processing}} in {{Support}} of {{Evidence-based Medicine}}},
  booktitle = {Findings of the {{Association}} for {{Computational Linguistics}}: {{ACL}} 2025},
  author = {Xu, Zihan and Ma, Haotian and Ding, Yihao and Zhang, Gongbo and Weng, Chunhua and Peng, Yifan},
  editor = {Che, Wanxiang and Nabende, Joyce and Shutova, Ekaterina and Pilehvar, Mohammad Taher},
  year = 2025,
  month = jul,
  pages = {21421--21443},
  publisher = {Association for Computational Linguistics},
  address = {Vienna, Austria},
  doi = {10.18653/v1/2025.findings-acl.1103},
  urldate = {2026-07-30},
  isbn = {979-8-89176-256-5},
}

@article{wangFoundationModelHumanAI2025a,
  title = {A Foundation Model for Human-{{AI}} Collaboration in Medical Literature Mining},
  author = {Wang, Zifeng and Cao, Lang and Jin, Qiao and Chan, Joey and Wan, Nicholas and Afzali, Behdad and Cho, Hyun-Jin and Choi, Chang-In and Emamverdi, Mehdi and Gill, Manjot K. and Kim, Sun-Hyung and Li, Yijia and Liu, Yi and Luo, Yiming and Ong, Hanley and Rousseau, Justin F. and Sheikh, Irfan and Wei, Jenny J. and Xu, Ziyang and Zallek, Christopher M. and Kim, Kyungsang and Peng, Yifan and Lu, Zhiyong and Sun, Jimeng},
  year = 2025,
  month = sep,
  journal = {Nature Communications},
  volume = {16},
  number = {1},
  pages = {8361},
  publisher = {Nature Publishing Group},
  issn = {2041-1723},
  doi = {10.1038/s41467-025-62058-5},
  urldate = {2026-07-30},
  copyright = {2025 The Author(s)},
  langid = {english},
}

@article{solliniHumanResearchersAre2025,
  title = {Human Researchers Are Superior to Large Language Models in Writing a Medical Systematic Review in a Comparative Multitask Assessment},
  author = {Sollini, Martina and Pini, Cristiano and Lazar, Alexandra and Gelardi, Fabrizia and Ninatti, Gaia and Bauckneht, Matteo and Chiti, Arturo and Kirienko, Margarita},
  year = 2025,
  month = dec,
  journal = {Scientific Reports},
  volume = {16},
  number = {1},
  pages = {173},
  publisher = {Nature Publishing Group},
  issn = {2045-2322},
  doi = {10.1038/s41598-025-28993-5},
  urldate = {2026-07-30},
  copyright = {2025 The Author(s)},
  langid = {english},
}

@article{soaresMakingScienceComputable2024,
  title = {Making {{Science Computable Using Evidence-Based Medicine}} on {{Fast Healthcare Interoperability Resources}}: {{Standards Development Project}}},
  shorttitle = {Making {{Science Computable Using Evidence-Based Medicine}} on {{Fast Healthcare Interoperability Resources}}},
  author = {Soares, Andrey and Schilling, Lisa M. and Richardson, Joshua and Kommadi, Bhagvan and Subbian, Vignesh and Dehnbostel, Joanne and Shahin, Khalid and Robinson, Karen A. and Afzal, Muhammad and Lehmann, Harold P. and Kunnamo, Ilkka and Alper, Brian S.},
  year = 2024,
  month = jun,
  journal = {Journal of Medical Internet Research},
  volume = {26},
  number = {1},
  pages = {e54265},
  publisher = {JMIR Publications Inc., Toronto, Canada},
  doi = {10.2196/54265},
  urldate = {2026-07-30},
  langid = {english},
}

@incollection{mckenzieSynthesizingPresentingFindings2024,
author    = {McKenzie, Joanne E. and Brennan, Sue E.},
title     = {Synthesizing and Presenting Findings Using Other Methods},
booktitle = {Cochrane Handbook for Systematic Reviews of Interventions},
edition   = {Version 6.5},
editor    = {Higgins, Julian P. T. and Thomas, James and Chandler, Jacqueline
and Cumpston, Miranda and Li, Tianjing and Page, Matthew J.
and Welch, Vivian A.},
publisher = {Cochrane},
year      = {2024},
chapter   = {12},
url       = {https://www.cochrane.org/authors/handbooks-and-manuals/handbook/current/chapter-12#section-12-3},
urldate   = {2026-07-30}
}

@inproceedings{mozgaiAcceleratingScopingReviews2024,
  title = {Accelerating {{Scoping Reviews}}: {{A Case Study}} in the {{User-Centered Design}} of an {{AI-Enabled Interdisciplinary Research Tool}}},
  shorttitle = {Accelerating {{Scoping Reviews}}},
  booktitle = {Extended {{Abstracts}} of the {{CHI Conference}} on {{Human Factors}} in {{Computing Systems}}},
  author = {Mozgai, Sharon A and Kaurloto, Cari and Winn, Jade G and Leeds, Andrew and Beland, Sarah and Sookiassian, Arman and Hartholt, Arno},
  year = 2024,
  month = may,
  series = {{{CHI EA}} '24},
  pages = {1--8},
  publisher = {Association for Computing Machinery},
  address = {New York, NY, USA},
  doi = {10.1145/3613905.3637110},
  urldate = {2026-07-30},
  isbn = {979-8-4007-0331-7},
}

@article{schlanderHowMuchDoes2021,
  title = {How {{Much Does It Cost}} to {{Research}} and {{Develop}} a {{New Drug}}? {{A Systematic Review}} and {{Assessment}}},
  shorttitle = {How {{Much Does It Cost}} to {{Research}} and {{Develop}} a {{New Drug}}?},
  author = {Schlander, Michael and {Hernandez-Villafuerte}, Karla and Cheng, Chih-Yuan and {Mestre-Ferrandiz}, Jorge and Baumann, Michael},
  year = 2021,
  month = nov,
  journal = {PharmacoEconomics},
  volume = {39},
  number = {11},
  pages = {1243--1269},
  issn = {1179-2027},
  doi = {10.1007/s40273-021-01065-y},
  urldate = {2026-03-11},
  langid = {english},
}

@article{singhDrugDiscoveryDevelopment2023,
  title = {Drug Discovery and Development: Introduction to the General Public and Patient Groups},
  shorttitle = {Drug Discovery and Development},
  author = {Singh, Natesh and Vayer, Philippe and Tanwar, Shivalika and Poyet, Jean-Luc and Tsaioun, Katya and Villoutreix, Bruno O.},
  year = 2023,
  month = may,
  journal = {Frontiers in Drug Discovery},
  volume = {3},
  publisher = {Frontiers},
  issn = {2674-0338},
  doi = {10.3389/fddsv.2023.1201419},
  urldate = {2026-02-16},
  langid = {english},
}

@misc{donevaLargeScaleAssessmentAnimaltoHuman2026,
  title = {Large-{{Scale Assessment}} of {{Animal-to-Human Drug Translation Using Natural Language Processing}}},
  author = {Doneva, Simona E. and Ellendorff, Tilia R. and Schneider, Gerold and Held, Leonhard and von Wyl, Viktor and Simpson, T. Ian and Sick, Beate and Ineichen, Benjamin V.},
  year = 2026,
  month = may,
  primaryclass = {New Results},
  pages = {2026.05.20.726540},
  publisher = {bioRxiv},
  issn = {2692-8205},
  doi = {10.64898/2026.05.20.726540},
  urldate = {2026-08-01},
  archiveprefix = {bioRxiv},
  chapter = {New Results},
  copyright = {\copyright{} 2026, Posted by openRxiv. This pre-print is available under a Creative Commons License (Attribution-NonCommercial 4.0 International), CC BY-NC 4.0, as described at http://creativecommons.org/licenses/by-nc/4.0/},
  langid = {english},
}

@article{hogueAssociationStatisticalMethodology2025,
  title = {{{Association}} of {{Statistical Methodology}} and {{Design}} in {{Preclinical Animal Studies With Successful Translation Into Clinical Phase}} 2 {{Trials}}},
  author = {Hogue, Olivia and Zelinsky, Megan and Sonneborn, Claire and Anantasagar, Thwisha and Tumma, Sanjana and Genuario, Isabella and Prabhakar, Tanvi and Smith, Emily and Adebulu, Jumoke and Salvaterra, Marissa and Pasadyn, Felicia L. and Dolansky, Mary A. and Obuchowski, Nancy A. and Baker, Kenneth and {Barnholtz-Sloan}, Jill S.},
  year = 2025,
  month = nov,
  journal = {Neurology},
  volume = {105},
  number = {9},
  pages = {e214250},
  issn = {0028-3878, 1526-632X},
  doi = {10.1212/WNL.0000000000214250},
  urldate = {2025-12-03},
  langid = {english},
}

@article{zeissEstablishedPatternsAnimal2017,
  title = {{{Established}} Patterns of Animal Study Design Undermine Translation of Disease-Modifying Therapies for {{Parkinson}}'s Disease},
  author = {Zeiss, Caroline J. and Allore, Heather G. and Beck, Amanda P.},
  year = {2017},
  journal = {PLOS ONE},
  volume = {12},
  number = {2},
  pages = {e0171790},
  publisher = {Public Library of Science},
  issn = {1932-6203},
  doi = {10.1371/journal.pone.0171790},
  urldate = {2025-03-05},
  langid = {english},
}

@article{wangPICOEntityExtraction2022a,
  title = {{{PICO}} Entity Extraction for Preclinical Animal Literature},
  author = {Wang, Qianying and Liao, Jing and Lapata, Mirella and Macleod, Malcolm},
  year = 2022,
  month = sep,
  journal = {Systematic Reviews},
  volume = {11},
  number = {1},
  pages = {209},
  issn = {2046-4053},
  doi = {10.1186/s13643-022-02074-4},
  urldate = {2026-02-26},
  langid = {english},

}

@inproceedings{choeSupportingNoviceResearchers2024,
  title = {Supporting {{Novice Researchers}} to {{Write Literature Review}} Using {{Language Models}}},
  booktitle = {Extended {{Abstracts}} of the {{CHI Conference}} on {{Human Factors}} in {{Computing Systems}}},
  author = {Choe, Kiroong and Park, Seokhyeon and Jung, Seokweon and Kim, Hyeok and Yang, Ji Won and Hong, Hwajung and Seo, Jinwook},
  year = 2024,
  month = may,
  series = {{{CHI EA}} '24},
  pages = {1--9},
  publisher = {Association for Computing Machinery},
  address = {New York, NY, USA},
  doi = {10.1145/3613905.3650787},
  urldate = {2026-07-30},
  isbn = {979-8-4007-0331-7},
}

@article{bahorDevelopmentUptakeOnline2021c,
  title = {Development and Uptake of an Online Systematic Review Platform: The Early Years of the {{CAMARADES Systematic Review Facility}} ({{SyRF}})},
  shorttitle = {Development and Uptake of an Online Systematic Review Platform},
  author = {Bahor, Zsanett and Liao, Jing and Currie, Gillian and Ayder, Can and Macleod, Malcolm and McCann, Sarah K and {Bannach-Brown}, Alexandra and Wever, Kimberley and Soliman, Nadia and Wang, Qianying and {Doran-Constant}, Lee and Young, Laurie and Sena, Emily S and Sena, Chris},
  year = 2021,
  month = mar,
  journal = {BMJ Open Science},
  volume = {5},
  number = {1},
  pages = {e100103},
  issn = {2398-8703},
  doi = {10.1136/bmjos-2020-100103},
  urldate = {2026-08-01},
  pmcid = {PMC8647599},
  pmid = {35047698}
}

@article{thurzoRevisitingRoleReview2025,
  title = {Revisiting the {{Role}} of {{Review Articles}} in the {{Age}} of {{AI-Agents}}: {{Integrating AI-Reasoning}} and {{AI-Synthesis Reshaping}} the {{Future}} of {{Scientific Publishing}}},
  shorttitle = {Revisiting the {{Role}} of {{Review Articles}} in the {{Age}} of {{AI-Agents}}},
  author = {Thurzo, Andrej and Varga, Ivan},
  year = 2025,
  month = apr,
  journal = {Bratislava Medical Journal},
  volume = {126},
  number = {4},
  pages = {381--393},
  issn = {1336-0345},
  doi = {10.1007/s44411-025-00106-8},
  urldate = {2026-08-01},
  langid = {english},
}

\appendix

\section{Interview Protocol}
\label{app:interview-protocol}
Our semi-structured interviews included the following guide questions. When appropriate, follow-up questions were used to encourage participants to elaborate on their responses, for instance to probe deeper into process details or recall motivations for a particular decision. 
\subsection*{Introduction}

Thank you very much for taking the time to speak with me today. My name is
\textbf{[NAME]}, and I am a \textbf{[ROLE]} within the team.

The purpose of this session is to learn more about your experience working
with published literature to inform animal and translational research. If I
ask ``why'' frequently, it is simply to ensure that I understand your
experiences correctly.

There are no right or wrong answers. You are free to stop the interview at
any time. If you are not comfortable answering a question, please let us
know, and we will skip it.

The session should last approximately 60 minutes and consists of
three parts:

\begin{itemize}[leftmargin=*, nosep]
    \item 5 minutes: warm-up and introduction questions;
    \item 25 minutes: problem exploration and brainstorming;
    \item 30 minutes: tool demonstration and feedback.
\end{itemize}

With your permission, we will record the session for research purposes only.

\medskip

\noindent\textbf{Do you have any questions before we begin?}

\subsection*{Warm-Up}

To begin, I would like to learn a little about your background.

\begin{enumerate}[leftmargin=*, label=\arabic*.]
    \item What is your current career level?
    \item What do you do in your current role?
    \item Do you currently work with animals, or have you worked with animals
    in the past?
    \item When you hear the term ``translational research,'' what does it mean
    in the context of your own work?
    \item Have you previously conducted or contributed to a systematic review?
\end{enumerate}

\subsection*{Topic 1: Current Approaches, Expectations, and Prior Knowledge}

\subsubsection*{Literature-Analysis Workflow}

\begin{enumerate}[leftmargin=*, label=\arabic*.]
    \item Can you walk me through how you typically search for and explore
    biomedical literature?
    \item Which biomedical databases do you use when analysing literature?
    \item Do you use any additional data resources beyond published articles?
    \item Do you use any AI tools to support literature-based research?
    \item Approximately how much time do you spend on this task each week?
    \item When you identify several studies on a topic, how do you evaluate
    which ones to examine further?
    \item What information do you usually look for in a study?
    \item How do you organize or keep track of the studies and results you find?
\end{enumerate}

\subsubsection*{Challenges and Information Needs}

\begin{enumerate}[leftmargin=*, label=\arabic*., resume]
    \item Which parts of this process are the most time-consuming or difficult?
    \item What are the main challenges you face when working with published
    literature?
    \item What information is particularly difficult to locate within papers?
    \item What is most difficult to compare across studies?
\end{enumerate}

\subsubsection*{Connecting Animal and Human Evidence}

\begin{enumerate}[leftmargin=*, label=\arabic*., resume]
    \item Do you ever need to relate animal studies to human or clinical studies?
    \item Can you walk me through how you make connections between animal and
    human evidence?
    \item What challenges do you encounter when making these connections through
    the literature?
    \item How do you judge whether an animal model is sufficiently relevant to
    inform human research or clinical decisions?
    \item In your view, how can animal or preclinical studies inform
    translational or clinical decision-making?
    \item Conversely, how can human or clinical studies inform earlier-stage
    animal or preclinical research?
\end{enumerate}

\subsubsection*{The Ideal Tool}

\begin{enumerate}[leftmargin=*, label=\arabic*., resume]
    \item If you had the perfect tool for analysing biomedical literature, which
    part of the process would you most want it to make easier?
    \item What would such a tool need to show in order to support translational
    research?
    \item Is there anything you wish existed but does not currently exist?
    \item What would success look like to you in this context?
\end{enumerate}

\subsection*{Topic 2: System Demonstration and Evaluation}

\subsubsection*{Interviewer Instructions}

Provide a high-level description of the available data and views. Then allow
the participant approximately three minutes to explore the system while
thinking aloud.

\medskip

\noindent

\subsubsection*{Initial Reactions}

\begin{enumerate}[leftmargin=*, label=\arabic*.]
    \item What do you think this tool is doing?
    \item What stands out to you?
    \item Is anything confusing or unclear at first glance?
    \item What do you think about how the information is presented?
    \item Would you prefer the information in another format?
\end{enumerate}

\subsubsection*{Potential Use}

\begin{enumerate}[leftmargin=*, label=\arabic*., resume]
    \item How might you use this output in your work?
    \item At what stage of your workflow would this tool be useful?
    \item Would it replace or complement something you currently do?
    \item Returning to our earlier discussion of translation, how might this tool
    support connections between animal and human evidence?
    \item Who else might benefit from this tool?
\end{enumerate}

\subsubsection*{Missing Information and Desired Functionality}

\begin{enumerate}[leftmargin=*, label=\arabic*., resume]
    \item Is anything missing or unclear in the output?
    \item What additional information would you expect the tool to provide?
    \item Are there any features or functions you would expect?
\end{enumerate}

\subsubsection*{Trust, Risks, and Limitations}

\begin{enumerate}[leftmargin=*, label=\arabic*., resume]
    \item Is there anything you would not trust or would want to verify?
    \item Would you use this tool in your work?
    \item What concerns or risks would you have?
    \item In which situations would you not rely on the tool?
\end{enumerate}

\subsubsection*{Potential Contribution to the 3Rs}

\begin{enumerate}[leftmargin=*, label=\arabic*., resume]
    \item At what stage of planning an animal or translational study could a tool
    like this support better decisions related to Replacement, Reduction, or
    Refinement?
\end{enumerate}

\noindent\textit{Possible prompts:}

\begin{itemize}[leftmargin=*]
    \item identifying existing evidence;
    \item highlighting weak translational assumptions;
    \item identifying possible refinements or alternatives;
    \item informing whether a new animal study is justified.
\end{itemize}

\subsection*{Closing Questions}

\begin{enumerate}[leftmargin=*, label=\arabic*.]
    \item What should I have asked you but did not?
    \item Is there anything else you would like to discuss?
    \item Do you have any questions for me?
\end{enumerate}

\medskip

Thank you again for taking the time to speak with me and for sharing your
experiences.

\section{Desired Evidence-Extraction Characteristics}
\label{app:extraction-characteristics}

Participants identified a broad set of study characteristics that evidence-support tools should extract and structure to facilitate comparison, synthesis, and assessment across studies. Table~\ref{tab:desired-extracted-characteristics} summarizes these characteristics and the participants who raised them.

\begin{table*}[th]
    \centering
    \caption{Study characteristics participants wanted support tools to extract and structure.}
    \label{tab:desired-extracted-characteristics}
    \small
    \begin{tabularx}{\textwidth}{
        >{\raggedright\arraybackslash}p{0.18\textwidth}
        >{\raggedright\arraybackslash}X
        >{\raggedright\arraybackslash}p{0.18\textwidth}}
        \toprule
        \textbf{Category}
        & \textbf{Desired information}
        & \textbf{Participants} \\
        \midrule

        Publication metadata
        &
        Article title; authors or main authors; research group or institution
        responsible for the study
        &
        P05, P09
        \\

        Study provenance
        &
        Country or study location; clinical-trial sponsor; preregistration
        status; registry link; and the location of extracted information in the
        article or supplementary materials
        &
        P02, P05, P09, P11
        \\

        Research topic
        &
        Disease; disease phenotype; behaviour or behavioural phenotype; brain
        region; and cognitive or biological concepts such as memory or sensory
        processing
        &
        P05, P07, P09, P10
        \\

        Animal model and population
        &
        Species; strain; exact genetic model; sex; age or life stage, including
        young, adolescent, adult, and aged animals; and experimental-group
        composition
        &
        P01, P02, P03, P04, P05, P07, P08, P10, P13
        \\

        Experimental structure
        &
        Individual experiments within a publication; number of groups; number
        of animals per group; experimental timeline; and the days or stages at
        which animals entered the experiment
        &
        P01, P06, P10
        \\

        Intervention
        &
        Drug or intervention; dose; treatment duration; route and method of
        administration; surgical, chemical, or genetic induction method;
        biological target or receptor; and mode or mechanism of action
        &
        P01, P02, P05, P06, P07, P08, P10, P11, P13
        \\

        Assays and procedures
        &
        Assay type; specific behavioural task; antibody used for histology;
        implantation or surgical procedure; and other assay-level methodological
        details
        &
        P02, P05, P07, P08, P10, P13
        \\

        Quantitative outcomes
        &
        Numerical readouts; effect sizes; power analyses; dose--response
        information; and EC50 values
        &
        P05, P06, P13
        \\

        Negative and unsuccessful evidence
        &
        Studies or experiments with negative results; failed experiments; and
        the experimental conditions associated with unsuccessful outcomes
        &
        P02, P05, P06
        \\

        Housing and animal welfare
        &
        Housing conditions; administration methods relevant to animal burden;
        anesthesia and analgesia; monitoring procedures; side effects;
        animal-welfare outcomes; and refinement measures
        &
        P02, P08, P10, P13
        \\

        Clinical study population
        &
        Participant sex, age, ethnicity, disease subtype, and other population
        characteristics required to assess the relevance of clinical evidence
        &
        P05, P11, P12
        \\

        \bottomrule
    \end{tabularx}
\end{table*}

\clearpage
\section{Desired Search and Filtering Functions}
\label{app:search-functions}

Participants also described functions for narrowing, prioritizing, and navigating large evidence sets according to their specific questions and contexts. Table~\ref{tab:desired-search-functions} summarizes the search, filtering, and prioritization capabilities they requested.

\begin{table*}[th]
    \centering
    \caption{Search, filtering, and prioritization functions requested by participants.}
    \label{tab:desired-search-functions}
    \small
    \begin{tabularx}{\textwidth}{
        >{\raggedright\arraybackslash}p{0.27\textwidth}
        >{\raggedright\arraybackslash}X
        >{\raggedright\arraybackslash}p{0.18\textwidth}}
        \toprule
        \textbf{Function}
        & \textbf{Description}
        & \textbf{Participants} \\
        \midrule

        Species-specific filtering
        &
        Restrict results to a selected species, such as dogs or rats, to reduce
        irrelevant results when comparing animal models
        &
        P02, P04
        \\

        Filtering by animal characteristics
        &
        Filter by strain, exact model, sex, age or life stage, and other
        population characteristics
        &
        P02, P05, P08, P10
        \\

        Filtering by intervention and assay
        &
        Filter by drug, dose, target, mechanism of action, assay type,
        behavioural task, or experimental procedure
        &
        P02, P05, P06, P08, P11
        \\

        Filtering by disease and phenotype
        &
        Search by disease, disease phenotype, behaviour, behavioural phenotype,
        or related clinical and experimental characteristics
        &
        P02, P05, P06, P07, P09, P10
        \\

        Searching by brain region or concept
        &
        Retrieve studies concerning a selected brain region or broader
        biological or cognitive concept, such as prefrontal cortex, memory, or
        sensory processing
        &
        P05, P07, P09
        \\

        Field-specific keyword search
        &
        Search for keywords within the title, abstract, full text, or selected
        article sections
        &
        P02, P05, P09
        \\

        Geographic filtering
        &
        Restrict or compare evidence by country or jurisdiction where geographic
        or regulatory context is relevant
        &
        P02, P09, P11
        \\

        Personalized filtering
        &
        Adapt available filters and retrieved results to the user's role,
        research question, and experimental context
        &
        P04, P05, P09, P10
        \\

        Relevance-based ranking
        &
        Rank studies according to their alignment with the user's research
        context and explain why particular studies may be relevant
        &
        P05, P06, P09
        \\

        Influence-based prioritization
        &
        Prioritize large result sets using transparent but imperfect signals
        such as citation counts and journal reputation
        &
        P06, P09
        \\

        Quality-related assessment
        &
        Display configurable indicators related to study quality or
        ``goodness'' without reducing appraisal to a single definitive score
        &
        P02, P05, P06, P09, P10, P13
        \\

        \bottomrule
    \end{tabularx}
\end{table*}

\clearpage
\section{Design Criteria for AI-Assisted Evidence Systems}
\label{app:design-criteria}

Beyond specific extraction and search functions, participants' accounts pointed to broader requirements for how AI-assisted evidence systems should support biomedical evidence work. Table~\ref{tab:design-criteria-rq3} synthesizes these needs into design criteria spanning evidence retrieval, representation, synthesis, translational and 3R decision support, and system trustworthiness.

\begin{table*}[t]
\caption{Design criteria for future AI-assisted biomedical evidence systems,
derived from participants' information needs and use contexts.}
\label{tab:design-criteria-rq3}

\footnotesize
\setlength{\tabcolsep}{4pt}
\renewcommand{\arraystretch}{1.12}

\begin{tabularx}{\textwidth}{
    @{}
    L{0.17\textwidth}
    L{0.22\textwidth}
    Y
    L{0.14\textwidth}
    @{}
}
\toprule
\textbf{Design criterion} &
\textbf{User need} &
\textbf{Actionable requirement} &
\textbf{Participants} \\
\midrule

\rowcolor{gray!12}
\multicolumn{4}{@{}l}{\textbf{Retrieval and exploration}} \\

Context-sensitive interaction &
Evidence adapted to the user's role, question, and research context &
Use information about the user, research question, model, and experimental
purpose to retrieve and explain relevant studies. Support iterative follow-up
questions and retain context across interactions. &
P05, P06 \\

\addlinespace[2pt]

Persistent evidence workspace &
Continuity across searches and research sessions &
Allow users to store papers, searches, prompts, notes, and extracted
information in a topic-specific workspace and cross-reference evidence across
studies. &
P10 \\

\addlinespace[2pt]

Semantic and relational search &
Access to related evidence beyond exact keyword matches &
Map synonyms, disease terminology, brand and substance names, mechanisms,
phenotypes, and outcomes. Support exploration of relationships among models,
targets, interventions, readouts, and clinical findings. &
P02, P05, P07, P09, P11, P12, P13 \\

\midrule

\rowcolor{gray!12}
\multicolumn{4}{@{}l}{\textbf{Evidence representation and assessment}} \\

Experiment-level structuring &
Direct access to the experiments reported within publications &
Separate individual models, groups, doses, assays, interventions, and
outcomes. Extract key methodological, quantitative, and welfare information
from articles and supplementary materials. &
P01, P02, P05, P08, P10, P13 \\

\addlinespace[2pt]

Cross-study comparison &
Structured examination of differences and conflicting findings &
Align studies by model, intervention, endpoint, timing, assay, and design.
Expose methodological differences, missing information, negative findings,
and uncertainty rather than collapsing evidence into a single score. &
P01, P05, P06, P09, P12, P13 \\

\addlinespace[2pt]

Human-controlled assessment &
Automation without loss of contextual scientific judgment &
Allow users to inspect and correct extracted information, adjust assessment
criteria, and reject system interpretations. Avoid definitive automated
judgments of study quality or relevance. &
P01, P02, P05, P06, P07, P12 \\

\midrule

\rowcolor{gray!12}
\multicolumn{4}{@{}l}{\textbf{Synthesis and output generation}} \\

Traceable evidence synthesis &
Faster production of reviews, reports, and licence-related text &
Generate drafts from a user-selected evidence collection and link each claim
to the supporting publication, experiment, and source passage. &
P05, P09, P10, P12 \\

\midrule

\rowcolor{gray!12}
\multicolumn{4}{@{}l}{\textbf{Translational and 3R decision support}} \\

Animal--human evidence linking &
Connections across animal, human, clinical, and regulatory evidence &
Link studies through mapped terminology, mechanisms, corresponding outcomes,
disease stages, and interventions. Make the evidence supporting each proposed
connection visible. &
P02, P03, P08, P09, P10, P11, P13 \\

\addlinespace[2pt]

Translational comparison &
Understanding patterns of translational success and failure &
Compare successful and unsuccessful translation and expose relevant
differences in models, procedures, interventions, and measured outcomes,
without presenting associations as definitive causal explanations. &
P09, P13 \\

\addlinespace[2pt]

3R and welfare support &
Evidence relevant to whether and how animal studies proceed &
For a proposed model or procedure, retrieve comparable experiments and
surface reported analgesia, welfare measures, refinements, less-burdensome
procedures, replacement approaches, and potentially unnecessary repetition. &
P02, P04, P05, P08, P10, P11, P13 \\

\addlinespace[2pt]

Licence review support &
Evidence for preparing and assessing proposed animal experiments &
Compare proposed experiments or licence applications with existing evidence
and highlight prior studies, relevant refinements, established replacements,
and weaknesses in model justification for expert review. &
P02, P05, P10, P11 \\

\midrule

\rowcolor{gray!12}
\multicolumn{4}{@{}l}{\textbf{Trust and reliability}} \\

Inspectable provenance &
Ability to verify extracted and generated information &
Link extracted values, comparisons, and generated claims to the exact
supporting passage and original publication. Clearly separate retrieved
evidence from system-generated interpretation. &
P05, P07, P12, P13 \\

\addlinespace[2pt]

Coverage and uncertainty awareness &
Understanding what the system searched and what may be missing &
Disclose searched databases and document types, indicate likely gaps, and
distinguish information that was unreported, not retrieved, or inferred by
the system. &
P01, P04, P05, P09, P10, P11, P12, P13 \\

\bottomrule
\end{tabularx}
\end{table*}

\section{Qualitative Analysis Materials}
\label{app:qualitative-analysis}

This appendix provides additional examples of the qualitative analysis process described in Section~\ref{sec:methods}. The screenshots illustrate the transcript coding in QualCoder, evolving codebook, and collaborative theme development.

\begin{figure*}[t]
    \centering
    \includegraphics[width=\textwidth]{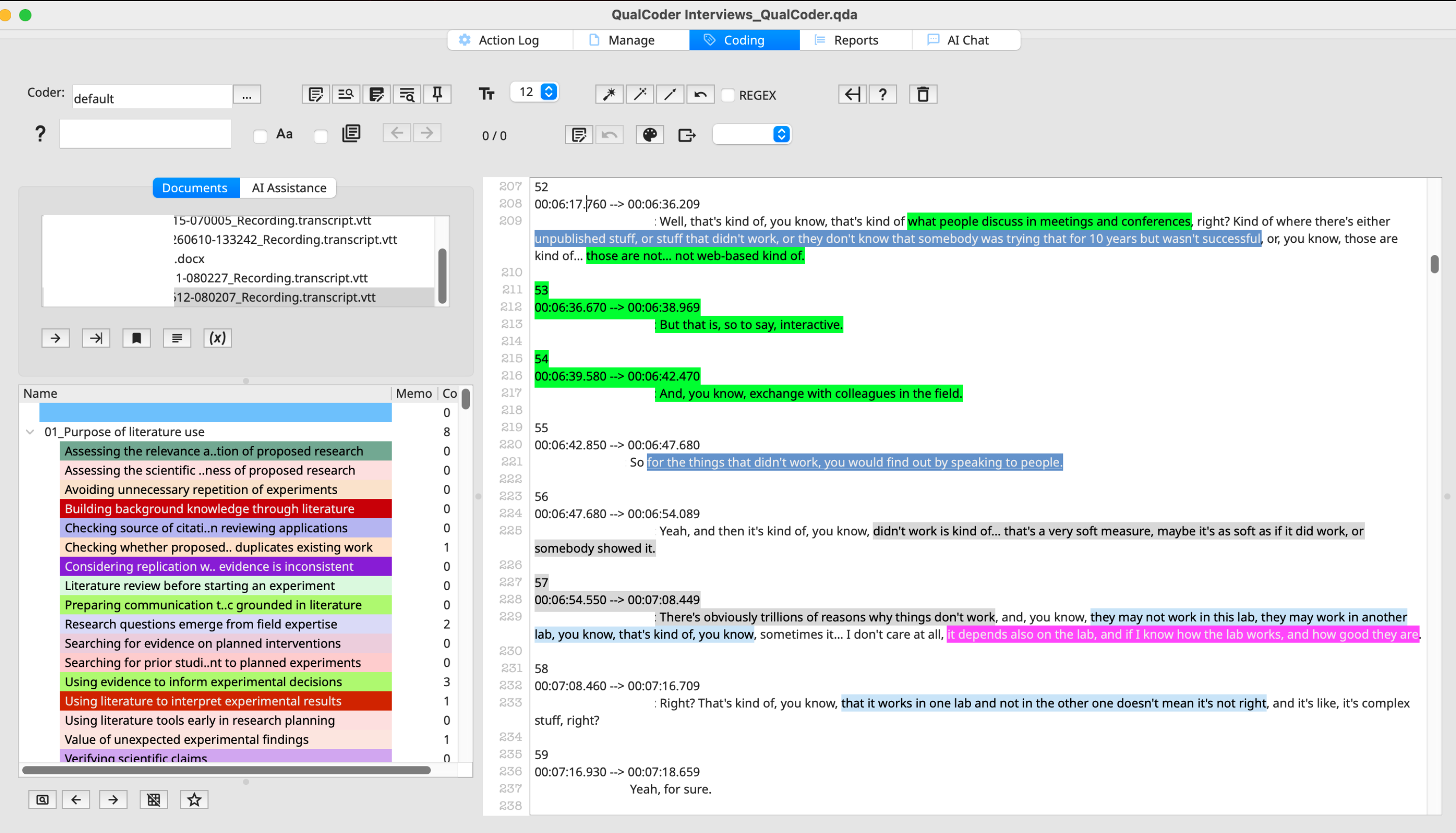}
    \caption{Example of transcript coding in QualCoder. Relevant transcript segments were assigned codes during the reflexive thematic analysis.}
    \Description{Screenshot of the QualCoder interface showing an interview transcript with coded text segments and the associated coding panel.}
    \label{fig:appendix-qualcoder}
\end{figure*}

\begin{figure*}[t]
    \centering
    \includegraphics[width=\textwidth]{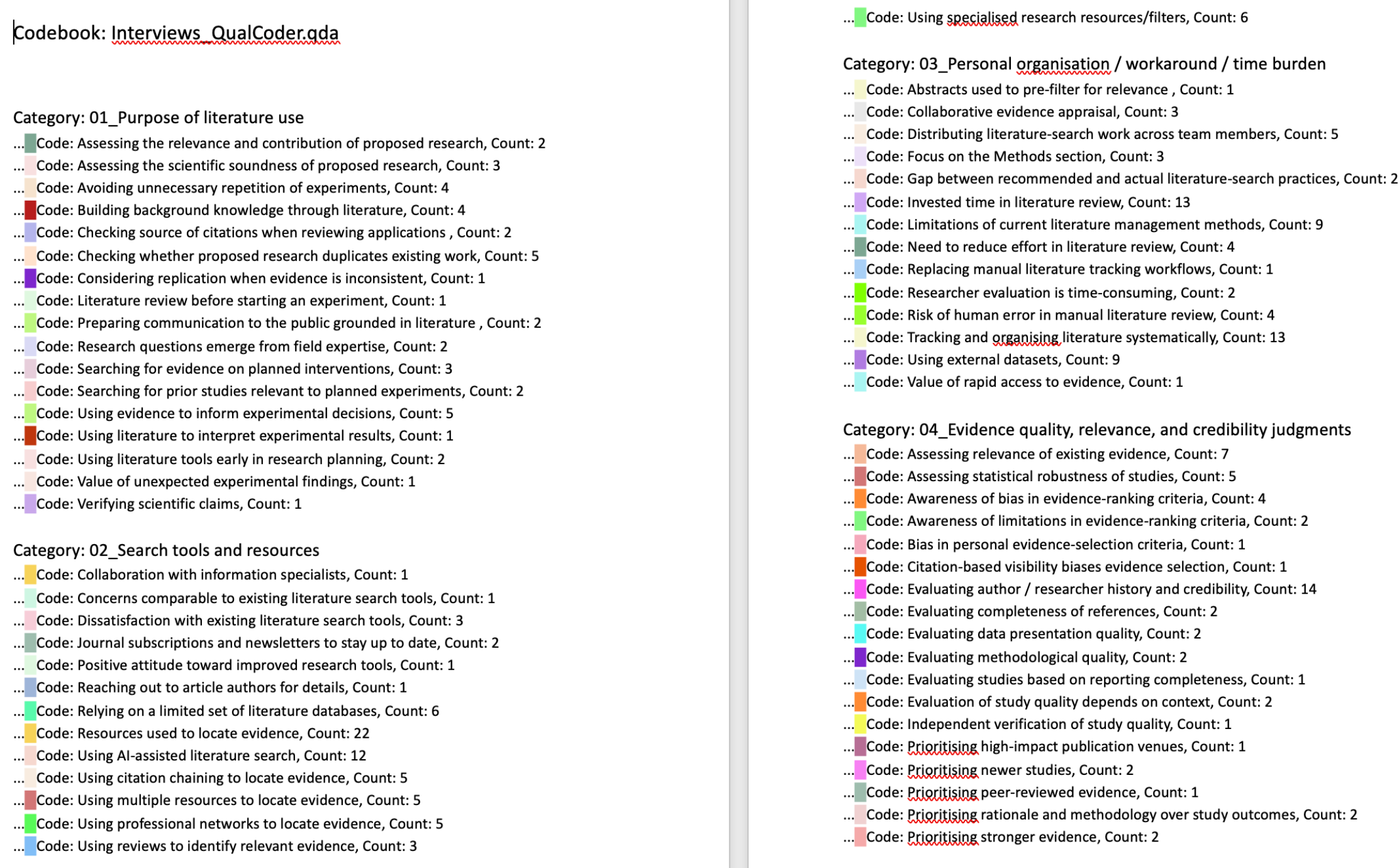}
    \caption{Excerpt from the qualitative codebook developed during the analysis. Codes were iteratively refined as the researchers coded and compared interview transcripts.}
    \Description{Screenshot showing an excerpt of the qualitative codebook,
    including code categories and individual codes used during interview
    analysis.}
    \label{fig:appendix-codebook}
\end{figure*}

\begin{figure*}[t]
    \centering
    \includegraphics[width=\textwidth]{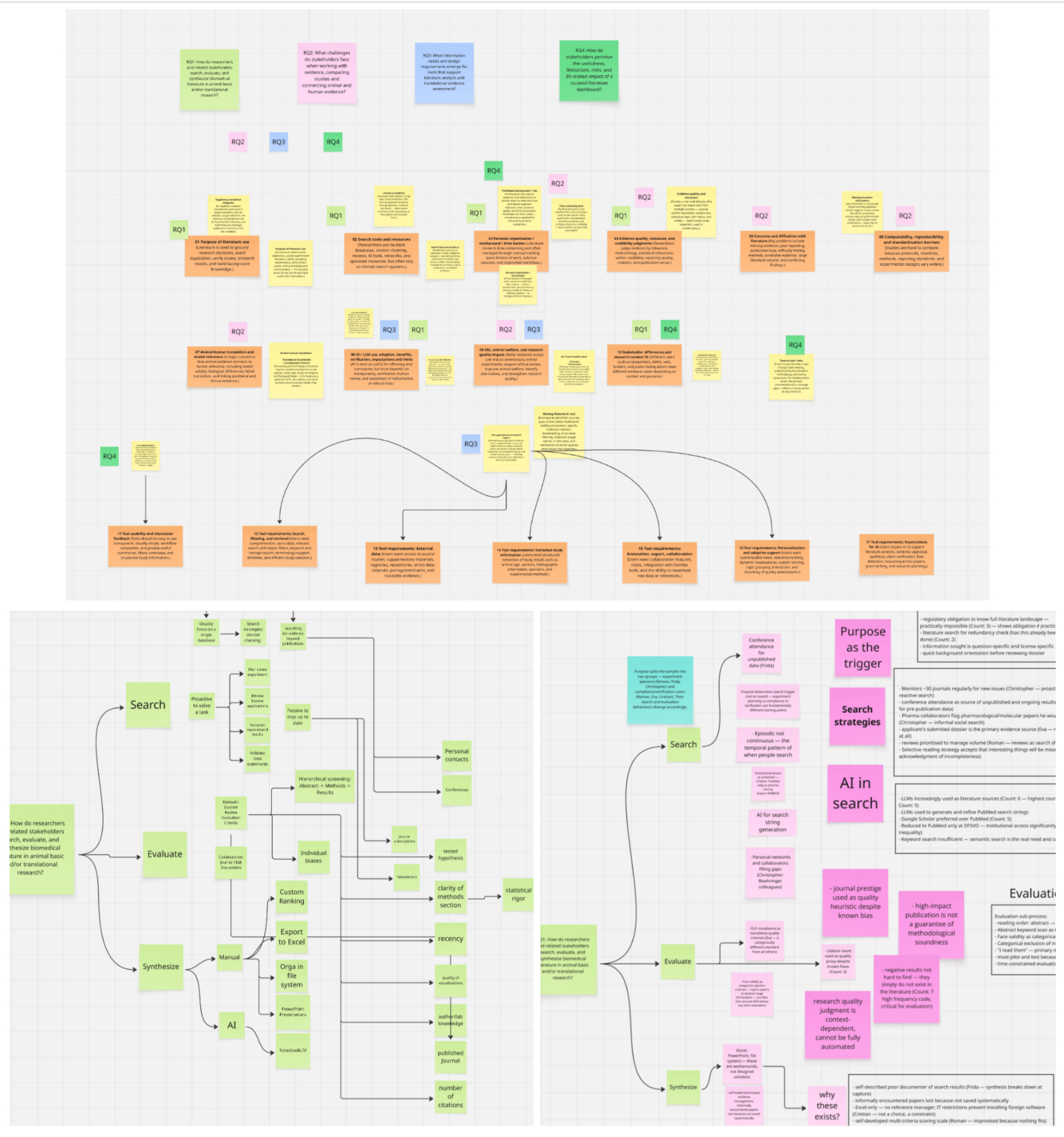}
    \caption{Example of collaborative theme development in Miro. Codes and coded observations were iteratively grouped, reorganized, and discussed to develop candidate themes across the interview dataset.}
    \Description{Screenshot of a Miro board used by the research team to
    organize qualitative codes and develop candidate themes.}
    \label{fig:appendix-miro}
\end{figure*}


\end{document}